\documentclass{article} 
\usepackage{emulateapj} 
\usepackage{apjfonts} 

\usepackage{natbib}
\usepackage{psfig}
\slugcomment{\hbox to 6.5in{Accepted for publication in 
{\em The Astrophysical Journal} \hfil           \hfil
\hbox{LBNL-41801}}}

\begin{document}
\tightenlines



\title{Measurements of $\Omega$ and $\Lambda$ from 42 High-Redshift Supernovae}
\vspace{0.1in}
\author{
S.~Perlmutter\altaffilmark{1},
G.~Aldering,
G.~Goldhaber\altaffilmark{1},
R.A.~Knop,
P.~Nugent,\\
P.~G.~Castro\altaffilmark{2},
S.~Deustua,
S.~Fabbro\altaffilmark{3},
A.~Goobar\altaffilmark{4},\\
D.~E.~Groom,
I.~M.~Hook\altaffilmark{5},
A.~G.~Kim\altaffilmark{1,6},
M.~Y.~Kim,
J.~C.~Lee\altaffilmark{7},\\
N.~J.~Nunes\altaffilmark{2},
R.~Pain\altaffilmark{3},
C.~R.~Pennypacker\altaffilmark{8},
R.~Quimby\\
}
\vspace{0.025in}

\affil{ Institute for Nuclear and Particle Astrophysics, \\
E.~O. Lawrence Berkeley National Laboratory,
Berkeley, California 94720.}
\vspace{0.07in}

\author{C.~Lidman}
\vspace{-0.02in}
\affil{ European Southern Observatory, La Silla, Chile.}

\author{R.~S.~Ellis, M. Irwin, R.~G.~McMahon}
\affil{ Institute of Astronomy, Cambridge, United Kingdom.}

\author{P.~Ruiz-Lapuente}
\affil{ Department of Astronomy, University of Barcelona, Barcelona, Spain.}

\author{N.~Walton}
\affil{ Isaac Newton Group, La Palma, Spain.}

\author{B.~Schaefer}
\affil{ Department of Astronomy, Yale University, New Haven, Connecticut.}

\author{B.~J.~Boyle}
\affil{ Anglo-Australian Observatory, Sydney, Australia.}

\author{A.~V.~Filippenko, T.~Matheson}
\affil{ Department of Astronomy, University of California, Berkeley, CA.}

\author{A.~S.~Fruchter, N.~Panagia\altaffilmark{9}}
\affil{ Space Telescope Science Institute, Baltimore, Maryland.}

\author{H.~J.~M.~Newberg}
\affil{ Fermi National Laboratory, Batavia, Illinois.}

\author{W.~J.~Couch}
\affil{ University of New South Wales, Sydney, Australia.}

\vspace{0.07in}
\author{(The Supernova Cosmology Project)}
\vspace{0.2in}

\altaffiltext{1}{Center for Particle Astrophysics, U.C. Berkeley, California.}
\altaffiltext{2}{Instituto Superior T\'{e}cnico, Lisbon, Portugal.}
\altaffiltext{3}{LPNHE, CNRS-IN2P3 \& University of Paris VI \& VII,
Paris, France.}
\altaffiltext{4}{Department of Physics, University of Stockholm, Stockholm, Sweden.}
\altaffiltext{5}{European Southern Observatory, Munich, Germany.}
\altaffiltext{6}{PCC, CNRS-IN2P3 \& Coll\`ege de France, Paris, France.}
\altaffiltext{7}{Institute of Astronomy, Cambridge, United Kingdom.}
\altaffiltext{8}{Space Sciences Laboratory, U.C. Berkeley, California.}
\altaffiltext{9}{Space Sciences Department, European Space Agency.}

\begin{abstract}

We report measurements of the mass density, $\Omega_{\rm M}$, and
cosmological-constant energy density, $\Omega_\Lambda$, of the universe
based on the analysis of 42 Type Ia supernovae discovered by
the Supernova Cosmology Project.  The magnitude-redshift data for
these supernovae, at redshifts between 0.18 and 0.83,
are fit jointly with a set of supernovae
from the Cal\'{a}n/Tololo Supernova Survey,
at redshifts below 0.1, to yield values for the cosmological parameters. All
supernova peak magnitudes are standardized using a SN~Ia
lightcurve width-luminosity relation.
The measurement yields a joint probability distribution of the
cosmological parameters that is approximated
by the relation $0.8 \,\Omega_{\rm M}- 0.6\,\Omega_\Lambda \approx
-0.2 \pm 0.1$ in the region of interest ($\Omega_{\rm M} \lesssim 1.5$).
For a flat
($\Omega_{\rm M}+\Omega_\Lambda = 1$) cosmology we find
$\Omega_{\rm M}^{\rm flat}  = 0.28^{+0.09}_{-0.08}$
(1$\sigma$ statistical)
$^{+0.05}_{-0.04}$ (identified systematics).  The data are strongly
inconsistent with a $\Lambda = 0$ flat cosmology, the simplest
inflationary universe model.
An open, $\Lambda = 0$ cosmology also
does not fit the data well:
the data indicate that the
cosmological constant is non-zero and positive, with
a confidence of $P(\Lambda > 0) = 99$\%,
including the identified systematic uncertainties.  The best-fit
age of the universe relative to the Hubble time is
$t_0^{\rm flat}=14.9^{+1.4}_{-1.1}\,(0.63/h)$ Gyr for a flat cosmology.
The size of our sample allows us to perform a variety of statistical
tests to check for possible systematic errors and biases. We find no
significant differences in either the host reddening distribution or
Malmquist bias between the low-redshift Cal\'{a}n/Tololo sample and our
high-redshift sample. Excluding those
few supernovae which are outliers in color excess
or fit residual does not significantly change the results.
The conclusions are also robust
whether or not a width-luminosity
relation is used to standardize the supernova peak magnitudes.
We discuss, and constrain where possible, hypothetical
alternatives to a cosmological constant.
\vspace{0.08in}

\end{abstract}


\section{Introduction}

Since the earliest studies of supernovae, it has been suggested that
these luminous events might be used as standard candles for
cosmological measurements \cite[]{baade38}.  At closer distances they
could be used to measure the Hubble constant, if an
absolute distance scale or magnitude scale could be established, while
at higher redshifts they could determine the deceleration parameter
\cite[]{tamm79, col79}.  The Hubble constant measurement became a
realistic possibility in the 1980's, when the more homogeneous subclass
of Type Ia supernovae (SNe~Ia) was identified
\cite[see][]{branch98}.  Attempts to measure the
deceleration parameter, however, were stymied for lack of high-redshift
supernovae.  Even after an impressive multi-year effort by
\cite{nn88}, it was only possible to follow one SN~Ia, at $z = 0.31$,
discovered 18 days past
its peak brightness.

The Supernova Cosmology Project was started in 1988 to address this
problem.  The primary goal of the project is the determination of the
cosmological parameters of the universe using the magnitude-redshift
relation of Type Ia supernovae.  In particular, \cite{omol_95} showed
the possibility of separating the relative contributions of the mass
density, $\Omega_{\rm M}$, and the cosmological constant, $\Lambda$, to
changes in the expansion rate by studying supernovae at a range of
redshifts.  The Project developed techniques, including
instrumentation, analysis, and observing strategies, that make it
possible to systematically study high-redshift supernovae
\cite[]{perl96}.  As of March 1998, more than 75 Type Ia supernovae at
redshifts $z = 0.18$--0.86 have been discovered and studied by the
Supernova Cosmology Project \cite[]{perliau95a, perliau96a, perliau97a,
perliau97b, perliau97c, perliau97d, perliau98a}.

A first presentation of analysis techniques, identification of possible
sources of statistical and systematic errors, and first results based
on seven of these supernovae at redshifts $z\sim 0.4$ were given in
\citeauthor{perl97} (\citeyear{perl97}; hereafter referred to as
``P97'').  These first results yielded a confidence region that was
suggestive of a flat, $\Lambda = 0$ universe, but with a large range of
uncertainty.  \cite{97ap} added a $z=0.83$ SN~Ia to this sample, with
observations from the Hubble Space Telescope and Keck 10-meter
telescope, providing the first demonstration of the method of separating
$\Omega_{\rm M}$ and $\Lambda$ contributions.
This analysis offered preliminary evidence for a
low-mass-density universe with a best-fit value of $\Omega_{\rm M} =
0.2 \pm 0.4$, assuming $\Lambda = 0$.  Independent work by
\cite{98hst_hiz}, based on three supernovae at $z \sim 0.5$ and one at
$z = 0.97$, also suggested a low mass density, with best-fit
$\Omega_{\rm M} = -0.1 \pm 0.5$ for $\Lambda = 0$.

\cite{perl97a} presented a preliminary analysis of 33 additional
high-redshift supernovae, which gave a confidence region indicating an
accelerating universe, and barely including a low-mass $\Lambda=0$
cosmology.  Recent independent work of \cite{riess16sne},
based on 10 high-redshift supernovae added to
the \citeauthor{98hst_hiz} set, reached the same conclusion.
Here we report on the complete analysis of 42 supernovae from the
Supernova Cosmology Project,
including the reanalysis of our previously reported supernovae with
improved calibration data and improved photometric and spectroscopic
SN~Ia templates.

\section{Basic Data and Procedures}

The new supernovae in this sample of 42 were all discovered while still
brightening, using the Cerro Tololo 4-meter telescope with the $2048^2$-pixel
prime-focus CCD camera or the $4 \times 2048^2$-pixel Big Throughput Camera
\cite[]{bernsteintyson}.  The supernovae were followed with photometry
over the peak of their lightcurves, and approximately two-to-three months
further ($\sim$40--60 days restframe) using the CTIO 4-m, WIYN 3.6-m, ESO
3.6-m, INT 2.5-m, and WHT 4.2-m telescopes.  (SN 1997ap and other
1998 supernovae have also been followed with HST photometry.)
The supernova redshifts and
spectral identifications were obtained using the Keck~I and II 10-m
telescopes with LRIS \cite[]{oke95} and the ESO 3.6-m telescope.  The photometry
coverage was most complete in Kron-Cousins $R$-band, with Kron-Cousins
$I$-band photometry coverage ranging from two or three points near peak
to relatively complete coverage paralleling the $R$-band observations.

Almost all of the new supernovae were observed spectroscopically.  The
confidence of the Type Ia classifications based on these spectra taken
together with the observed lightcurves, ranged from ``definite'' (when
Si II features were visible) to ``likely'' (when the features were
consistent with Type Ia, and inconsistent with most other types).  The
lower confidence identifications were primarily due to host-galaxy
contamination of the spectra.  Fewer than 10\% of the original sample
of supernova candidates from which these SNe~Ia were selected were
confirmed to be non-Type Ia, i.e., being active galactic nuclei
or belonging to another SN
subclass; almost all of these non-SNe~Ia could also have been identified
by their lightcurves and/or position far from the SN~Ia Hubble line.
Whenever possible, the redshifts were measured from the
narrow host-galaxy lines, rather than the broader supernova lines.
The lightcurves and several spectra are shown in \citeauthor{97ap}
(\citeyear{perl97}, \citeyear{perl97a}, \citeyear{97ap}); complete
catalogs and detailed discussions of the photometry and spectroscopy
for these supernovae will be presented in forthcoming papers.

The photometric reduction and the analyses of the lightcurves followed
the procedures described in P97.
The supernovae were observed with the Kron-Cousins filter that
best matched the restframe $B$ and $V$ filters at the
supernova's redshift, and any remaining mismatch of wavelength coverage
was corrected by calculating the expected photometric
difference---the ``cross-filter $K$-correction''---using template SN~Ia
spectra, as in \cite{kim_kcorr96}.  We have now recalculated these
$K$ corrections \cite[see][]{nuge_kcorr98}, using improved
template spectra, based on
an extensive database of low-redshift SN~Ia spectra recently made
available from the Cal\'{a}n/Tololo survey \cite[]{phillips98}.
Where available, IUE
and HST spectra \cite[]{iuesne, kir92a} were also
added to the SN~Ia spectra, including those published previously
\cite[1972E, 1981B, 1986G, 1990N, 1991T, 1992A, and 1994D
in:][]{kir72e, bra81b, phil86g, jeffetal92, meik94d, patat94d}.  In
\cite{nuge_kcorr98} we show that the $K$-corrections can be calculated
accurately for a given day on the supernova lightcurve, and for
a given supernova lightcurve width, from the color of the supernova
on that day.
(Such a calculation of $K$ correction based on supernova color will
also automatically account for any modification of the $K$ correction
due to reddening of the supernova; see \citeauthor{nuge_kcorr98}
\citeyear{nuge_kcorr98}.  In the case of insignificant reddening the
SN~Ia template color curves can be used.)  We find that
these calculations are robust to mis-estimations of the lightcurve
width or day on the lightcurve, giving results correct to within 0.01
mag for lightcurve width errors of $\pm 25$\% or lightcurve phase
errors of $\pm 5$ days even at redshifts where filter matching is the worst.
Given small additional uncertainties in the colors of supernovae,
we take an overall systematic uncertainty of 0.02 magnitudes
for the $K$ correction.

The improved $K$-corrections have been recalculated for all the
supernovae used in this paper, including those previously analyzed and
published.  Several of the low-redshift supernovae from the
Cal\'{a}n/Tololo survey have relatively large changes (as much as 0.1
magnitudes) at times in their
$K$-corrected lightcurves.  (These and other low-redshift supernovae
with new $K$-corrections are used by several independent groups in
constructing SN~Ia lightcurve templates, so the templates must be
updated accordingly.) The $K$-corrections for several of the
high-redshift supernovae analyzed in P97 have also changed by
small amounts at the lightcurve peak ($\Delta K(t\!=\!0)\lesssim 0.02$ mag)
and somewhat larger amounts by 20 days past peak
($\Delta K(t\!=\!20)\lesssim 0.1$ mag); this
primarily affects the measurement of the restframe lightcurve width.
These $K$-correction changes balance out
among the P97 supernovae, so
the final results for these supernovae do not change significantly.
(As we discuss below, however, the much larger current dataset does affect the
interpretation of these results.).

As in P97, the peak magnitudes have been corrected for the lightcurve
width-luminosity relation of SNe~Ia:
\begin{equation}
m_B^{\rm corr} = m_B + \Delta_{\rm corr}(s),
\end{equation}
where the correction term $\Delta_{\rm corr}$
is a simple monotonic function of the ``stretch factor,'' $s$,
that stretches or contracts the time
axis of a template SN~Ia lightcurve
to best fit the observed lightcurve
for each supernova
(see Perlmutter et al. 1995a, 1997e; \citeauthor{kim_stretch98} \citeyear{kim_stretch98}; \citeauthor{gold_dilate98} \citeyear{gold_dilate98}; 
and cf. \citeauthor{phillipsdeltam15} \citeyear{phillipsdeltam15};
\citeauthor{rpk95} \citeyear{rpk95}, \citeyear{rpk96}).
A similar relation corrects the $V$ band lightcurve, with the same
stretch factor in both bands.
For the supernovae discussed in this paper, the template must
be time-dilated by a factor $1+z$ before fitting to the
observed lightcurves to account
for the cosmological lengthening of the supernova timescale
\cite[]{goldhaberaigua,leibundgutdilation,riess_age97}.
P97 calculated $\Delta_{\rm corr}(s)$ by translating from
$s$ to $\Delta m_{15}$ (both describing the timescale of the supernova
event) and then using the relation
between $\Delta m_{15}$ and luminosity as determined by \cite{hametal95}.
The lightcurves of the Cal\'{a}n/Tololo supernovae have since been
published, and we have directly fit each lightcurve with the stretched template
method to determine its stretch factor $s$.
In this paper, for the light-curve width-luminosity relation,
we therefore directly use the functional form
\begin{equation}
\Delta_{\rm corr}(s) = \alpha (s-1)
\end{equation}
and determine $\alpha$ simultaneously with our determination of
the cosmological parameters.   With this functional form,
the supernova peak apparent magnitudes are thus all ``corrected'' as they
would appear if the supernovae had the lightcurve width of the template, $s=1$.

We use analysis procedures that are designed to be as similar as possible
for the low- and high-redshift datasets.  Occasionally,
this requires not using all of the data available at low redshift, when
the corresponding data are not accessible at high redshift.
For example, the low-redshift supernova lightcurves can often be followed
with photometry for
many months with high signal-to-noise ratios, whereas the high-redshift
supernova observations are generally only practical for approximately
60 restframe days past maximum light.  This
period is also the phase of the low-redshift SN~Ia lightcurves that is
fit best by the stretched-template method, and best predicts the
luminosity of the supernova at maximum.  We therefore fit only this period for
the lightcurves of the low-redshift supernovae.
Similarly, at high redshift the restframe $B$-band photometry is usually much
more densely sampled in time than the restframe $V$-band data, so we use the
stretch factor that best fits the restframe $B$ band data for both low-
and high-redshift supernovae, even though at low-redshift the
$V$-band photometry is equally well sampled.

Each supernova peak magnitude was also
corrected for Galactic extinction, $A_R$, using the extinction law of
\cite{card89}, first using the color excess, $E(B\!-\!V)_{\rm SF\&D}$, at the
supernova's Galactic coordinates provided by \cite{schl_red98}\, and
then---for comparison---using the $E(B\!-\!V)_{\rm B\&H}$ value
provided by \cite{bh_red82, bh_red98}.
$A_R$ was calculated from $E(B\!-\!V)$
using a value of the total-to-selective extinction ratio,
${\cal R}_R \equiv A_R / E(B\!-\!V)$, specific to each supernova. These were calculated using the appropriate
redshifted supernova spectrum as it would appear through an $R$-band filter.
These values of ${\cal R}_R$ range from 2.56 at $z = 0$ to 4.88 at $z = 0.83$.
The observed supernova colors were similarly corrected for Galactic extinction.
Any extinction in the
supernova's host galaxy, or between galaxies, was not corrected for at
this stage, but will be analyzed separately in Section 4.

All the same corrections for width-luminosity relation,
$K$ corrections, and extinction
(but using ${\cal R}_B = 4.14$) were applied to the photometry of 18
low-redshift SNe~Ia ($z \le 0.1$) from the Cal\'{a}n/Tololo supernova
survey \cite[]{hametal96} that were discovered
earlier than five days after peak.
The lightcurves of these 18 supernovae have all been re-fit
since P97, using the more recently available photometry \cite[]{hametal96}
and our $K$~corrections.

Figures~\ref{hubdiaglog}~and~\ref{hubdiag}(a) show the Hubble diagram of effective rest-frame
$B$ magnitude corrected for the width-luminosity relation,
\begin{equation}
m_B^{\rm effective} = m_R + \Delta_{\rm corr} - K_{BR} - A_R
\end{equation}
as a function of redshift for the 42 Supernova Cosmology
Project high-redshift supernovae, along with the 18 Cal\'{a}n/Tololo
low-redshift supernovae. (Here, $K_{BR}$ is the cross-filter $K$ correction
from observed $R$ band to restframe $B$ band.)
Tables~\ref{scpdata} and \ref{hamuydata} give the
corresponding IAU names, redshifts, magnitudes, corrected magnitudes,
and their respective uncertainties.  As in P97,
the inner error bars in Figures~\ref{hubdiaglog}~and~\ref{hubdiag}
represent the photometric
uncertainty, while the outer error bars add in quadrature 0.17
magnitudes of intrinsic dispersion of SN~Ia magnitudes that remain
after applying the
width-luminosity correction.  For these plots, the slope of the
width-brightness relation was taken to be $\alpha = 0.6$,
the best-fit value of Fit C discussed below.
(Since both the low- and high-redshift supernova
light-curve widths are clustered rather closely around $s=1$, as
shown in Figure~\ref{stretchhists}, the exact
choice of $\alpha$ does not change the Hubble diagram
significantly.)  The theoretical curves for a universe
with no cosmological constant are shown as solid lines, for a range of
mass density, $\Omega_{\rm M} = 0, 1, 2$.  The dashed lines represent
alternative flat cosmologies, for which the total mass-energy density
$\Omega_{\rm M} + \Omega_\Lambda = 1$ (where $\Omega_\Lambda \equiv
\Lambda/3 H_0^2$).  The range of models shown are for ($\Omega_{\rm
M},\Omega_\Lambda$) = (0,1), (0.5,0.5), (1,0), which is covered by
the matching solid line, and (1.5, $-$0.5).

\section{Fits to $\Omega_{\rm M}$ and $\Omega_\Lambda$}

The combined low- and high-redshift supernova datasets of
Figure~\ref{hubdiaglog} are fit to the
Friedman-Robertson-Walker magnitude-redshift relation,
expressed as in P97:
\begin{eqnarray}
\lefteqn{m_B^{\rm effective} \equiv m_R +\alpha(s-1)  - K_{BR} -
A_R}\hspace*{1.2in} & &
\\ & & = {\cal M}_B + 5 \log {\cal D}_L(z;\Omega_{\rm
M},\Omega_\Lambda)\;,\nonumber \label{FRW}
\end{eqnarray}
where ${\cal D}_L \equiv H_0 d_L$ is the
``Hubble-constant-free'' luminosity distance and
${\cal M}_B \equiv M_B - 5 \log H_0 + 25$ is the
``Hubble-constant-free'' $B$-band absolute magnitude at maximum
of a SN~Ia with width $s=1$.
(These quantities are, respectively, calculated from theory or
fit from apparent magnitudes and redshifts, both without
any need for $H_0$.  The cosmological-parameter results are thus also completely
independent of $H_0$.)
The details of the fitting procedure as
presented in P97 were followed, except that both the low- and
high-redshift supernovae were fit simultaneously, so
that ${\cal M}_B$ and $\alpha$, the slope of the width-luminosity
relation, could also be fit in addition to
the cosmological parameters $\Omega_{\rm M}$ and
$\Omega_\Lambda$.  For most of the analyses in this paper,
${\cal M}_B$ and $\alpha$ are statistical ``nuisance'' parameters;
we calculate 2-dimensional confidence
regions and single-parameter uncertainties for the cosmological
parameters by integrating over these parameters,
i.e., ${\cal P}(\Omega_{\rm M},\Omega_\Lambda) =
\int \!\! \int {\cal P}(\Omega_{\rm M},\Omega_\Lambda,{\cal M}_B,\alpha)
d{\cal M}_B\;d\alpha$.

As in P97, the small correlations between the photometric uncertainties
of the high-redshift supernovae, due to shared calibration data, have been
accounted for by fitting with a correlation matrix of uncertainties.
(The correlation matrix is available at http://www-supernova.lbl.gov.)
The low-redshift supernova photometry is more likely to be uncorrelated in
its calibration since these supernovae were not discovered in batches.
However, we take a 0.01 mag systematic uncertainty in the comparison of
the low-redshift $B$-band photometry and the high-redshift $R$-band
photometry.  The stretch-factor uncertainty
is propagated with a fixed width-luminosity slope
(taken from the low-redshift supernovae; cf. P97), and checked for
consistency after the fit.

We have compared the results of Bayesian and classical,
``frequentist,'' fitting
procedures.  For the Bayesian fits, we have assumed a ``prior''
probability distribution that
has zero probability for $\Omega_{\rm M} < 0$, but otherwise
uniform probability in the four parameters $\Omega_{\rm M}$,
$\Omega_\Lambda$, $\alpha$, and ${\cal M}_B$.  For the frequentist fits,
we have followed the classical statistical
procedures described by \cite{feld98}, to guarantee frequentist
coverage of our confidence regions in the physically allowed part of
parameter space.  Note that throughout the previous cosmology
literature, completely unconstrained fits have generally been used
that can (and do) lead to confidence regions that include the part of
parameter space with negative values for $\Omega_{\rm M}$.
The differences between the confidence regions that result from
Bayesian and classical analyses are small.  We present the
Bayesian confidence regions in the figures, since they are somewhat
more conservative, i.e. have larger confidence regions, in the
vicinity of particular interest near $\Lambda = 0$.

The residual dispersion in SN Ia peak magnitude after correcting
for the width-luminosity relation is small, about 0.17 magnitudes,
before applying any color-correction.  This was reported in Hamuy
et al. (1996) for the low-redshift Calan-Tololo supernovae, and
it is striking that
the same residual is most consistent with the current 42 high-redshift
supernovae (see Section 5).
It is not clear from the current datasets, however, whether
this dispersion is best modeled as a normal distribution (a Gaussian
in flux space) or a log-normal distribution (a Gaussian in magnitude
space).  We have therefore performed the fits two ways:
minimizing $\chi^2$ measured using either magnitude
residuals or flux residuals.  The results
are generally in excellent agreement, but since the magnitude fits
yield slightly larger confidence regions, we have again chosen this
more conservative alternative to report in this paper.

We have analyzed the total set of 60 low- plus high-redshift supernovae
in several ways, with the results of each fit presented as a row of
Table~\ref{fitresults}.  The most inclusive analyses are presented in the first
two rows: Fit~A is a fit to the entire dataset, while
Fit~B excludes two supernovae that are the most significant
outliers from the average lightcurve width, $s=1$, and
two of the remaining supernovae that are the largest residuals from Fit~A.
Figure~\ref{stretchhists}
shows that the remaining low- and high-redshift supernovae are
well matched in their lightcurve
width---the error-weighted means are
$\langle s \rangle_{\rm Hamuy} = 0.99 \pm 0.01$
and $\langle s \rangle_{\rm SCP} =
1.00 \pm 0.01$---making the results robust with respect to the width-luminosity-relation
correction (see Section 4.5).  Our primary analysis, Fit~C,
further excludes two supernovae
that are likely to be reddened, and is discussed in the following section.

Fits~A and B give very similar results.  Removing the two large-residual
supernovae from
Fit~A yields indistinguishable results, while Figure~\ref{confmulti}(a)
shows that the 68\% and 90\%
joint confidence regions for $\Omega_{\rm M}$ and $\Omega_\Lambda$
still change very little after also removing the two supernovae
with outlier lightcurve widths.
The best-fit mass-density in a flat
universe for Fit~A is, within a fraction
of the uncertainty, the same value as for Fit~B,
$\Omega_{\rm M}^{\rm flat} = 0.26^{+0.09}_{-0.08}$
(see Table~\ref{fitresults}).
The main difference between the fits is the
goodness-of-fit: the larger $\chi^2$ per
degree of freedom for Fit~A, $\chi^2_\nu= 1.76$,
indicates that the outlier
supernovae included in this fit are probably
not part of a Gaussian distribution
and thus will not be appropriately weighted in a $\chi^2$ fit.  The
$\chi^2$ per degree of freedom for Fit~B, $\chi^2_\nu = 1.16$,
is over 300 times more probable than that of fit A, and indicates that the
remaining 56 supernovae are a reasonable fit to the model, with no large
statistical errors remaining unaccounted for.

Of the two large-residual
supernovae excluded from the fits after Fit~A, one
is fainter than the best-fit prediction and one is
brighter.
The photometric color excess (see Section~4.1) for
the fainter supernova, SN 1997O,
has an uncertainty that is too large to
determine conclusively whether it is reddened.
The brighter supernova, SN 1994H, is one of the first seven
high-redshift supernovae originally analyzed in P97, and is one of the
few supernovae without a spectrum to confirm its classification as
a SN~Ia.  After re-analysis
with additional calibration data and improved $K$-corrections, it
remains the brightest outlier in the current sample, but it affects the
final cosmological fits much less as one of 42 supernovae, rather than
1 of 5 supernovae in the primary P97 analysis.

\vspace{.3in}
\section{Systematic Uncertainties and Cross-checks}

With our large sample of 42 high-redshift SNe,
it is not only possible to obtain good statistical uncertainties on
the measured parameters,
but also to quantify several possible sources of systematic uncertainties.
As discussed in P97, the primary approach is to examine subsets of
our data that will be affected to lesser extents by the systematic
uncertainty being considered.  The high-redshift sample is now
large enough that these subsets each contain enough supernovae to yield
results of high statistical significance.

\subsection{Extragalactic Extinction.}

\subsubsection{Color-Excess Distributions}

Although we have accounted for extinction due to our Galaxy, it is
still probable that some supernovae are dimmed by host
galaxy dust or intergalactic dust.  For a standard dust extinction law
\cite[]{card89} the color, $B\!-\!V$, of a supernova will become redder as the
amount of extinction, $A_B$, increases.
We thus can look for any extinction differences between the
low- and high-redshift supernovae by comparing their 
restframe colors.
Since there is a small dependence of intrinsic color on the lightcurve width,
supernova colors can only be compared for the same stretch factor;
for a more convenient analysis, we subtract out the intrinsic
colors, so that the remaining color excesses can be compared
simultaneously for all stretch factors.
To determine the restframe color excess $E(B\!-\!V)$
for each supernova, we fit the rest-frame $B$ and $V$ photometry to the
$B$ and $V$ SN~Ia lightcurve templates, with one of the fitting
parameters representing the magnitude difference between the two
bands at their respective peaks.  Note that these lightcurve peaks are
$\sim$2 days apart, so the resulting $B_{\rm max}\!-\!V_{\rm max}$ color
parameter, which is frequently used to describe supernova colors, is
not a color measurement on a particular day.  The difference of this
color parameter from the $B_{\rm max}\!-\!V_{\rm max}$ found for a sample
of low-redshift supernovae for the same lightcurve stretch-factor
\cite[]{tripp98,kim_stretch98, phil_bmv98} does yield the restframe
$E(B\!-\!V)$ color excess for the fitted supernova.

For the high-redshift
supernovae at $0.3 < z < 0.7$,
the matching $R$- and $I$-band measurements take the place
of the restframe $B$ and $V$ measurements and the fit $B$ and $V$
lightcurve templates are $K$-corrected from the appropriate matching
filters, e.g. $R(t) = B(t) + K_{BR}(t)$ \cite[]{kim_kcorr96,nuge_kcorr98}.
For the
three supernovae at $z > 0.75$, the observed $R\!-\!I$ corresponds more
closely to a restframe $U\!-\!B$ color than to a $B\!-\!V$ color,
so $E(B\!-\!V)$ is calculated from restframe
$E(U\!-\!B)$ using the extinction law of \cite{card89}.  Similarly, for the
two SNe~Ia at $z \sim 0.18$, $E(B\!-\!V)$ is calculated from
restframe $E(V\!-\!R)$.

Figure~\ref{bminv} shows the
color excess distributions for both the
low- and high-redshift supernovae, after removing the color excess
due to our Galaxy.  Six high-redshift supernovae are not shown on
this $E(B\!-\!V)$ plot, because six of the first seven
high-redshift supernovae
discovered were not observed in both $R$ and $I$ bands.
The color of one low-redshift supernova, SN 1992bc, is poorly
determined by the $V$-band template fit and has also been excluded.
Two supernovae in the
high-redshift sample are $>3\sigma$ red-and-faint outliers from
the mean in the
joint probability distribution of $E(B\!-\!V)$ color excess and
magnitude residual from Fit~B.  These two, SNe 1996cg and 1996cn
(shown in light shading in Figure~\ref{bminv}), are very likely reddened
supernovae.  To obtain a more robust fit of the cosmological parameters,
in Fit~C
we remove these supernovae from the sample.  As can be seen from the Fit-C
68\% confidence region of Figure~\ref{confmulti}(a),
these likely-reddened supernovae
do not significantly affect any of our results.  The main distribution
of 38 high-redshift
supernovae thus is barely affected by a few reddened events.
We find identical results if we exclude the six supernovae without
color measurements (Fit~G in Table~\ref{fitresults}).
We take Fit~C to be our primary analysis for this paper, and in
Figure~\ref{40conf}, we show a more extensive range of confidence
regions for this fit.

\vspace{.25in}
\subsubsection{Cross-checks on Extinction}

The color-excess distributions of the Fit~C dataset
(with the most significant
measurements highlighted by dark shading in
Figure~\ref{bminv}) show no significant difference between the low-
and high-redshift means.  The dashed curve drawn over the
high-redshift distribution of Figure~\ref{bminv} shows the expected
distribution if the low-redshift distribution
had the measurement uncertainties of the high-redshift supernovae
indicated by the dark shading. This shows
that the reddening distribution for the high-redshift SNe is
consistent with the reddening distribution for the low-redshift SNe,
within the measurement uncertainties.  The error-weighted
means of the low- and
high-redshift distributions are almost identical:
$\langle{E(B\!-\!V)}\rangle_{\rm Hamuy} \;=
0.033 \pm 0.014$ mag and $\langle{E(B\!-\!V)}\rangle_{\rm SCP}\; = 0.035 \pm
0.022$ mag.  We also find no significant
correlation between the color excess and the statistical
weight or redshift of the
supernovae within these two redshift ranges.

To test the effect of any remaining high-redshift reddening on the Fit~C
measurement of the cosmological parameters, we have constructed a
Fit~H-subset of the high-redshift supernovae that is intentional biased
to be bluer than the low-redshift sample.
We exclude the error-weighted reddest 25\% of the
high-redshift supernovae; this excludes 9 high-redshift supernovae
with the highest error-weighted $E(B\!-\!V)$.
We further exclude two supernovae that have large uncertainties
in $E(B\!-\!V)$ but are significantly faint in their residual from
Fit~C.  This is a somewhat conservative cut since it removes the
faintest of the high-redshift supernovae, but it does ensure that
the error-weighted $E(B\!-\!V)$ mean of the remaining
supernova subset is a good indicator of any reddening
that could affect the cosmological parameters.
The probability that the high-redshift subset of Fit~H
is redder in the mean than the low-redshift
supernovae is less than 5\%;
This subset is thus very unlikely
to be biased to fainter magnitudes by high-redshift reddening.  Even with non-standard, ``greyer'' 
dust that does not
cause as much reddening for the same amount of extinction,
a conservative estimate of
the probability that the high-redshift subset of Fit~H
is redder in the mean than the low-redshift
supernovae is still less than $\sim$17\%,
for any high-redshift value of ${\cal R}_B \equiv
A_B /E(B\!-\!V)$ less than twice the low-redshift value.
(These same confidence levels
are obtained whether using Gaussian statistics, assuming a normal
distribution of $E(B\!-\!V)$ measurements, or using bootstrap
resampling statistics, based on the observed distribution.)
The confidence regions of Figure~\ref{confmulti}(c) and the
$\Omega_{\rm M}^{\rm flat}$ results in Table~\ref{fitresults} show that the
cosmological parameters found for Fit~H differ by less than half of a standard
deviation from those for Fit~C.
We take the difference of these fits, 0.03 in $\Omega_{\rm M}^{\rm flat}$
(which corresponds to less than 0.025 in magnitudes)
as a $\sim$1$\sigma$ upper bound on the systematic uncertainty due to
extinction by dust that reddens.

Note that the modes of both distributions appear to be at zero
reddening, and similarly the medians of the distributions are quite
close to zero reddening:
$\langle{E(B\!-\!V)}\rangle_{\rm Hamuy}^{\rm median} \;= 0.01$ mag
and $\langle{E(B\!-\!V)}\rangle_{\rm SCP}^{\rm median}\; = 0.00$ mag.
This should be taken as suggestive
rather than conclusive since the zeropoint of the relationship between
true color and stretch is not tightly constrained by the current
low-redshift SN~Ia dataset.  This apparent strong clustering of SNe~Ia
about zero reddening has been noted in the past for low-redshift
supernova samples.  Proposed explanations have been given based on the
relative spatial distributions of the SNe~Ia and the dust:
Modeling by \cite{hatano97} of the expected extinction of SN~Ia disk
and bulge populations viewed at random orientations shows an extinction
distribution with a strong spiked peak near zero extinction along with a broad,
lower-probability wing to higher extinction.
This wing will be further suppressed by the observational
selection against more reddened SNe,
since they are dimmer.
(For a flux-limited survey this suppression
factor is $10^{- a_R \lbrack {\cal R}_B E(B\!-\!V) - \alpha (s-1)\rbrack}
\approx 10^{-1.6 E(B\!-\!V)}$, where
$a_R$ is the slope of the supernova number counts.)
We also note that the
high-redshift supernovae for which we have accurate measurements of
apparent separation between SN and host position (generally, those with
Hubble Space Telescope imaging) appear to be
relatively far from the host center, despite our high search
sensitivity to supernovae in front of the host galaxy core
(see \citeauthor{rate_96} \citeyear{rate_96} for search efficiency studies;
also cf. \citeauthor{wang97} \citeyear{wang97}).
If generally true for the entire sample, this would be consistent
with little extinction.

Our results, however, do not depend on the low- and high-redshift
color-excess distributions being consistent with zero reddening.
It is only important that the reddening
distributions for the low-redshift and high-redshift datasets are
statistically the same, and that there is no correlation between reddening
and statistical weight in the fit of the cosmological parameters.
With both of these conditions satisfied, we find that our measurement
of the cosmological parameters is unaffected (to within
the statistical error) by any small remaining extinction among the
supernovae in the two datasets.

\subsubsection{Analysis with Reddening Correction of \\
Individual Supernovae}

We have also performed fits using restframe $B$-band
magnitudes individually corrected for host galaxy extinction
using $A_B = {\cal R}_B E(B\!-\!V)$ (implicitly assuming
that the extragalactic extinction is all at the redshift of the
host galaxy).
As a direct comparison between the treatment of host galaxy extinction
described above and an alternative Bayesian
method \cite[]{rpk96}, we applied it to the 53 SNe~Ia with
color measurements in our Fit~C dataset.
We find that our cosmological parameter results are robust with respect
to this change, although
this method can introduce a bias into the extinction corrections,
and hence the cosmological parameters.
In brief, in this method the Gaussian extinction
probability distribution implied by the measured color-excess and its error
is multiplied by an assumed {\em a priori} probability distribution
(the Bayesian prior) for the intrinsic distribution of host
extinctions.  The most probable value of the resulting renormalized probability
distribution is taken as the extinction, and following Riess (private
communication) the second-moment is taken as the uncertainty.
For this analysis, we choose a conservative prior \cite[as given in][]{rpk96}
that does not {\em assume} that the supernovae are unextinguished,
but rather is somewhat broader than the
true extinction distribution where the majority of the previously observed
supernovae apparently suffer very little reddening.
(If one alternatively assumes that the current data's extinction distribution
is quite as narrow as that of previously observed supernovae, one can
choose a less conservative
but more realistic narrow prior probability distribution,
such as that of \cite{hatano97}.  This turns out to be quite similar to
our previous analysis in Section 4.1.1, since a distribution like that of \citeauthor{hatano97}
has zero extinction for most supernovae.)

This Bayesian method with a conservative prior
will only brighten supernovae, never make them fainter, since
it only affects the supernovae with redder measurements
than the zero-extinction $E(B\!-\!V)$ value,
leaving unchanged those measured to be bluer than this.
The resulting slight difference between the assumed and true
reddening distributions would make no difference in the cosmology
measurements if its size were the same at low and high redshifts.
However, since the uncertainties, $\sigma_{E(B\!-\!V)}^{{\rm high}-z}$,
in the high-redshift dataset $E(B\!-\!V)$ measurements are larger
on average than those of the low-redshift dataset,
$\sigma_{E(B\!-\!V)}^{{\rm low}-z}$, this method can
over-correct the high-redshift supernovae on average relative
to the low-redshift supernovae.
Fortunately, as shown in Appendix A,
even an extreme case with a true distribution
all at zero extinction and a conservative prior would introduce a bias in
extinction $A_B$ only of order 0.1 magnitudes at worst
for our current low- and high-redshift measurement uncertainties.
The results of Fit~E are shown in Table~\ref{fitresults} and as the
dashed contour in
Figure~\ref{confmulti}(d), where it can be seen that compared to Fit~C
this approach moves the best fit value much less than this,
and in the direction expected for this effect (indicated
by the arrows in Figure~\ref{confmulti}d).  The fact that
$\Omega_{\rm M}^{\rm flat}$ changes so little from Case~C, even
with the possible bias,
gives further confidence in the cosmological results.

We can eliminate any such small bias of this method
by assuming no Bayesian prior on the host-galaxy extinction, allowing
extinction corrections to be negative in the case of supernovae
measured to be bluer than the zero-extinction $E(B\!-\!V)$ value.
As expected, we recover the unbiased results within error, but with larger
uncertainties since the Bayesian prior also narrows
the error bars in the method of \cite{rpk96}.
However, there remains a
potential source of bias when correcting for reddening:
the effective ratio
of total to selective extinction, ${\cal R}_B$, could vary, for
several reasons.  First,
the extinction could be due to host galaxy dust at the supernova's redshift
or intergalactic dust
at lower redshifts, where it will redden the supernova less since it
is acting on a redshifted spectrum.  Second, ${\cal R}_B$ may be
sensitive to dust density, as indicated by
variations in the dust extinction laws between various
sight-lines in the Galaxy \cite[]{cc88, gc98}.
Changes in metallicity might be expected to be
a third possible cause of ${\cal R}_B$ evolution,
since metallicity is one dust-related quantity known to evolve with
redshift \cite[]{pettini97}, but fortunately it appears not to
significantly alter ${\cal R}_B$ as evidenced by the similarity of the optical
portions of the extinction curves of the Galaxy, the LMC, and the SMC \cite[]{pei92,gc98}.
Three-filter photometry of high-redshift supernovae
currently in progress with the Hubble Space Telescope will help test for
such differences in ${\cal R}_B$.

To avoid these sources of bias, we consider it important to
use and compare both analysis approaches: the
rejection of reddened supernovae and the correction of reddened
supernovae.  We do find consistency in the results calculated both ways.
The advantages of the analyses with
reddening corrections applied to individual supernovae
(with or without a Bayesian prior on host-galaxy extinction)
are outweighed by the disadvantages for our sample of high-redshift supernovae;
although, in principle, by applying reddening corrections the intrinsic
magnitude dispersion of SNe~Ia can be reduced
from an observed dispersion of 0.17 magnitudes to approximately 0.12
magnitudes, in practice the net
improvement for our sample
is not
significant since uncertainties in the color measurements
often dominate. We have therefore
chosen for our primary analysis to follow the
first procedure discussed above, removing the likely-reddened
supernovae (Fit~C) and then comparing color-excess means.
The systematic difference for Fit~H, which rejects
the reddest and the faintest high-redshift supernovae,
is already quite small,
and we avoid introducing additional actual and possible biases.
Of course, neither approach avoids biases if
${\cal R}_B$ at high redshift
is so large [$ >  2{\cal R}_B({z=0})$] that dust does not redden the
supernovae enough to be distinguished {\em and} this dust makes
more than a few supernovae faint.

\subsection{Malmquist Bias and other Luminosity Biases.}

In the fit of the cosmological parameters to the magnitude-redshift
relation, the low-redshift supernova magnitudes primarily
determine ${\cal M}_B$ and the width-luminosity slope $\alpha$, and then
the comparison with the high-redshift supernova magnitudes primarily
determines $\Omega_{\rm M}$ and $\Omega_\Lambda$.  Both low- and
high-redshift supernova samples can be biased towards
selecting the brighter tail of any
distribution in supernova {\it detection} magnitude for supernovae found
near the detection threshold of the search (classical Malmquist bias;
\citeauthor{malm24} \citeyear{malm24}, \citeyear{malm36}).
A width-luminosity relation fit
to such a biased population would have a slope that is slightly too
shallow and a zeropoint slightly too bright.
A second bias is also acting on the supernova samples, selecting
against supernovae on the
narrow-lightcurve side of the width-luminosity relation since such
supernovae are detectable for a shorter period of time.
Since this bias removes the narrowest/faintest supernova lightcurves
preferentially, it culls out the part of the width-brightness distribution
most subject to Malmquist bias, and moves the resulting best-fit slope
and zeropoint closer to their correct values.

If the Malmquist bias is the same in both datasets, then it is completely
absorbed by ${\cal M}_B$ and $\alpha$ and does not affect the cosmological
parameters. Thus, our principal concern is that there could be a
difference in the amount of bias between the low-redshift and
high-redshift samples.
Note that effects peculiar to photographic SNe searches, such as
saturation in galaxy cores, which might in principle select slightly different
SNe~Ia sub-populations should not be important in determining luminosity
bias because lightcurve stretch compensates for any such differences. Moreover,
Figure~\ref{stretchhists} shows that
the high-redshift SNe~Ia we have discovered have a stretch distribution
entirely consistent with those discovered in the Cal\'{a}n/Tololo search.

To estimate the Malmquist bias of the high-redshift-supernova sample,
we first determined the completeness of our high-redshift
searches as a function of magnitude,
through an extensive series of tests inserting artificial SNe into our images
\cite[see][]{rate_96}.
We find that roughly 30\% of our high-redshift
supernovae were detected within twice the SN~Ia intrinsic luminosity
dispersion of the 50\% completeness limit, where the above biases might
be important.  This is consistent with a simple model where the
supernova number counts follow a power-law slope of 0.4~mag$^{-1}$, similar
to that seen for comparably distant galaxies \cite[]{smail95}.
For a flux-limited survey of standard candles having
the lightcurve-width-corrected luminosity dispersion for SN~Ia of
$\sim$0.17~mag and this number-count power-law slope, we can calculate that the
classical Malmquist bias should be 0.03~mag \cite[see, e.g.,][for
a derivation of the classical Malmquist bias]{mihalasbinney}.  (Note that
this estimate is much smaller than the Malmquist bias affecting other
cosmological distance indicators, due to the much smaller intrinsic
luminosity dispersion of SNe~Ia.)
These high-redshift supernovae, however,
are typically detected before maximum, and their detection
magnitudes and peak magnitudes have a correlation coefficient of only
0.35, so the effects of classical Malmquist bias should be diluted.
Applying the formalism of \cite{willick94} we estimate that the
decorrelation between detection magnitude and peak magnitude
reduces the classical Malmquist bias in the high-redshift sample to only 0.01~mag.  The redshift and stretch distributions of the
high-redshift supernovae that are near the 50\%-completeness limit
track those of the overall high-redshift sample, again suggesting that
Malmquist biases are small for our dataset.

We cannot make an exactly parallel estimate of Malmquist bias for the
low-redshift-supernova sample, because we do not
have information for the Cal\'{a}n/Tololo dataset concerning the
number of supernovae found near the detection limit.
However, the amount of classical
Malmquist bias should be similar for the Cal\'{a}n/Tololo SNe since
the amount of bias is dominated by the intrinsic luminosity
dispersion of SNe~Ia, which we find to be the same for the low-redshift
and high-redshift samples (see Section 5). Figure~\ref{stretchhists} shows
that the stretch distributions for
the high-redshift and low-redshift samples are very similar,
so that the compensating
effects of stretch-bias should also be similar in the two datasets.
The major source of difference in the bias is expected to be due to the close
correlation between the detection magnitude and the peak magnitude
for the low-redshift supernova search, since this search tended not
to find the supernovae as early before peak as the high-redshift
search. In addition, the number-counts at low-redshift should be
somewhat steeper \cite[]{maddox90}. We thus expect the
Cal\'{a}n/Tololo SNe to have a bias closer to that obtained
by direct application of the the classical Malmquist bias
formula, 0.04~mag.
One might also expect ``inhomogeneous Malmquist bias'' to be more
important for the low-redshift supernovae,
since in smaller volumes of space inhomogeneities
in the host galaxy distribution might by chance put more supernovae near
the detection limit than would be expected for a homogeneous distribution.
However, after averaging over all the Cal\'{a}n/Tololo supernova-search fields
the total low-redshift volume searched is large enough that we expect galaxy
count fluctuations of only $\sim$4\%, so the classical Malmquist bias
is still a good approximation.

We believe that both these low- and high-redshift biases may be
smaller, and even closer to each other, due to the mitigating effect of the
bias against detection of low-stretch supernovae, discussed above.
However, to be conservative, we take the classical Malmquist
bias of 0.04~mag for the low-redshift dataset, and the least biased value
of 0.01~mag for the high-redshift dataset, and consider systematic uncertainty
from this source to be the difference, 0.03 mag, in the direction of
low-redshift supernovae more biased than high-redshift.   In the other direction, i.e.
for high-redshift supernovae more biased than low-redshift, we consider
the extreme case of a fortuitously unbiased low-redshift sample, and take
the systematic uncertainty bound to be the 0.01 mag bias of the high-redshift
sample.  (In this direction any systematic error is less relevant to the
question of the existence of a cosmological constant.)

\subsection{Gravitational Lensing.}

As discussed in P97, the clumping of mass in the universe could leave
the line-of-sight to most of the supernovae under-dense, while
occasional supernovae may be seen through over-dense regions.  The
latter supernovae could be significantly brightened by gravitational
lensing, while the former supernovae would appear somewhat fainter.
With enough supernovae, this effect will average out (for inclusive
fits, such as Fit A, which include outliers), but the most
over-dense lines of sight may be so rare that a set of 42 supernovae
may only sample a slightly biased (fainter) set.  The probability
distribution of these amplifications and deamplifications has previously been
studied both analytically and by Monte Carlo simulations.  Given the
acceptance window of our supernova search, we can integrate the
probability distributions from these studies
to estimate the bias due to amplified or deamplified
supernovae that may be rejected as outliers.  This average (de)amplification
bias is less than 1\% at the redshifts of our
supernovae for simulations based on isothermal spheres the size of
typical galaxies \cite[]{hw97}, N-body simulations using realistic mass
power spectra \cite[]{wamb98}, and the analytic models of
\cite{frie96}.

It is also possible that the small-scale clumping of matter is more
extreme, e.g., if significant amounts of mass were in the form of
compact objects such as MACHOs.  This could lead to many supernova
sightlines that are not just under-dense, but nearly empty.  Once
again, only the very rare line of sight would have a compact object in
it, amplifying the supernova signal.  To first approximation, with 42
supernovae we would see only the nearly empty beams, and thus only
deamplifications.
The appropriate luminosity-distance formula in this case is not the
Friedmann-Robertson-Walker (FRW) formula but rather the ``partially filled
beam'' formula with a mass filling factor, $\eta \approx 0$
\cite[see][and references therein]{kant98}.
We present the results of the fit of our data (Fit~K) with this
luminosity-distance formula \cite[as calculated using the code of][]{angsiz96}
in Figure~\ref{40confeta}.
A more realistic limit on this
point-like mass density can be estimated, because we would expect such
point-like masses to collect into the gravitational
potential wells already marked by
galaxies and clusters. \cite{fhp98} estimate an upper limit of
$\Omega_{\rm M} < 0.25$ on the mass which is clustered like galaxies.  In
Figure~\ref{40confeta}, we also show the confidence region
from Fit~L, assuming that only the mass density
contribution up to $\Omega_{\rm M} = 0.25$ is point-like, with filling factor
$\eta = 0$, and that $\eta$ rises to 0.75 at $\Omega_{\rm M} = 1$.
We see that at low mass density, the
Friedman-Robertson-Walker fit is already very close to the
nearly empty-beam ($\eta \approx 0$) 
scenario, so the results are quite similar.  At high mass
density, the results diverge, although only minimally for Fit~L;
the best fit in a flat universe is
$\Omega_{\rm M}^{\rm flat}  = 0.34^{+0.10}_{-0.09}$.

\subsection{Supernova Evolution and Progenitor Environment Evolution}

The spectrum of a SN~Ia on any given point in its lightcurve reflects
the complex physical state of the supernova on that day: the
distribution, abundances, excitations, and velocities of the elements
that the photons encounter as they leave the expanding photosphere all
imprint on the spectra. So far, the high-redshift supernovae that have
been studied have lightcurve shapes just like those of low-redshift supernovae
\cite[see][]{gold_dilate98}, and their spectra show the same features
on the same day of the lightcurve as their low-redshift counterparts
having comparable lightcurve width. This is true all the way out to the
$z = 0.83$ limit of the current sample \cite[]{97ap}.  We take this as
a strong indication that the physical parameters of the supernova explosions
are not evolving significantly over this time span.

Theoretically, evolutionary effects might be caused by changes
in progenitor populations or environments.  For example,
lower metallicity and more massive
SN~Ia-progenitor binary systems should be found in younger stellar populations.
For the redshifts that we are considering, $z < 0.85$,
the change in average progenitor masses may be small
\cite[]{pilaraigua,pilarheidelberg}.
However, such progenitor mass differences or differences in
typical progenitor metallicity are expected to
lead to differences in the final C/O ratio in the exploding white
dwarf, and hence affect the energetics of the explosion.  The
primary concern here would be if this changed the zero-point of
the width-luminosity relation.
We can look for such changes by comparing lightcurve rise
times between low and high-redshift supernova samples, since this
is a sensitive indicator of explosion energetics.  Preliminary
indications suggest that no significant rise-time change is seen,
with an upper limit of 
$\lesssim$1 day for our sample (see forthcoming high-redshift
studies of Goldhaber et al. 1998 and Nugent et al. 1998, and low-redshift
bounds from \citeauthor{vacca94d} \citeyear{vacca94d},
\citeauthor{leib90n} \citeyear{leib90n}, and
\citeauthor{marvin89b} \citeyear{marvin89b}).  This tight a constraint on
rise-time change would theoretically limit the zero-point change
to less than $\sim$0.1 mag  \cite[see][]{nuge_tsquare,hof_q098}.

A change in typical C/O ratio can also affect the ignition density
of the explosion and the propagation characteristics of the
burning front.  Such changes would be expected to appear as differences
in lightcurve timescales before and after maximum \cite[]{hofkhoklc96}.  Preliminary
indications of consistency between such low- and high-redshift
lightcurve timescales suggest that
this is probably not a major effect for our supernova samples
(Goldhaber et al., 1998).

Changes in typical progenitor metallicity should also directly cause
some differences in SN~Ia spectral features \cite[]{hof_q098}.
Spectral differences big enough to affect the $B$ and
$V$-band lightcurves
\cite[see, for example, the extreme mixing models presented
in Figure~9 of][]{hof_q098}
should be clearly visible for the best signal-to-noise spectra we have
obtained for our distant supernovae, yet they are not seen
(\citeauthor{filippenkoscp} \citeyear{filippenkoscp};
Hook, Nugent, et al., \citeyear{hook98}).  The consistency of slopes in the lightcurve width-luminosity relation for the low-
and high-redshift supernovae can also constrain the possibility of
a strong metallicity effect of the type that \cite{hof_q098} describes.

An additional concern might be that even small changes in spectral
features with metallicity could in turn affect the calculations of
$K$ corrections and reddening corrections.
This effect, too, is very small,
less than 0.01 magnitudes, for photometric observations
of SNe~Ia conducted in the restframe $B$ or $V$ bands
\cite[see Figures~8 and 10 of][]{hof_q098}, as is the case
for almost all of our supernovae.
(Only two of our supernovae have primary observations that are sensitive to the
restframe $U$ band, where the magnitude can change by $\sim$0.05 magnitudes, and
these are the two supernovae with the lowest weights in our fits, as
shown by the error bars of Figures~\ref{hubdiag}. In general the $I$-band
observations, which are mostly sensitive to the restframe $B$ band,
provide the primary lightcurve at redshifts above 0.7.)

The above analyses constrain only the
effect of progenitor-environment evolution on
SN~Ia intrinsic luminosity; however, the extinction of the supernova light
could also be affected, if the amount or character of the dust evolves, e.g.
with host galaxy age.  In Section 4.1, we limited the size of this
extinction evolution for
dust that reddens, but evolution of ``grey''
dust grains larger than $\sim$0.1$\mu$m,
which would cause more color-neutral optical extinction, can evade these color
measurements.  The following two analysis approaches
can constrain both evolution effects, intrinsic SN~Ia luminosity evolution and
extinction evolution.  They take advantage of the fact that galaxy 
properties such as formation age, star-formation history, and metallicity are
not monotonic functions of redshift, so even the low-redshift SNe~Ia
are found in galaxies with a wide range of ages and metallicities.  It
is a shift in the {\em distribution} of 
relevant host-galaxy properties occurring
between $z \sim 0$ and $z \sim 0.5$ that could cause any evolutionary effects.

\vspace{.1in} {\em Width-Luminosity Relation Across Low-Redshift Environments.}
To the extent that low-redshift SNe~Ia arise from progenitors with a
range of metallicities and ages, the lightcurve
width-luminosity relation discovered
for these SNe can already account for these effects
\cite[cf.][1996]{hametal95}.  When corrected for the width-luminosity
relation, the peak magnitudes of low-redshift SNe~Ia exhibit a
very narrow magnitude dispersion about the Hubble line, with no
evidence of a significant progenitor-environment
difference in the residuals from this fit.
It therefore does not matter if the population of progenitors evolves
such that the measured lightcurve widths change, since the
width-luminosity relation apparently is able to correct for these changes.
It will be important to continue to study further nearby SNe~Ia to
test this conclusion with as wide a range of
host-galaxy ages and metallicities as possible.

\vspace{.1in} {\em Matching Low- and High-Redshift Environments.}
Galaxies with different morphological classifications result from
different evolutionary histories.
To the extent that galaxies with similar classifications have
similar histories, we can also check
for evolutionary effects by using supernovae
in our cosmology measurements with matching host galaxy classifications.
If the same cosmological results are found for each measurement
based on a subset of low- and high-redshift supernovae sharing
a given host-galaxy classification,
we can rule out many evolutionary scenarios.  In the simplest
such test, we compare the cosmological parameters measured from
low- and high-redshift elliptical host galaxies with those measured
from low- and high-redshift spiral host galaxies.  Without
high-resolution host-galaxy images for most of our high-redshift sample,
we currently can only approximate this test for the smaller number
of supernovae for which the host-galaxy spectrum gives a strong
indication of galaxy classification.  The resulting sets of 9
elliptical-host and 8 spiral-host high-redshift supernovae
are matched to the 4 elliptical-host and 10 spiral-host low-redshift
supernovae \cite[based on the morphological classifications listed
in][and excluding two with SB0 hosts]{hametal96}.  We find no
significant change in
the best-fit cosmology for the elliptical host-galaxy subset (with
both the low- and high-redshift subsets about one sigma brighter than
the mean of the full sets), and a
small ($<$1$\sigma$) shift lower in $\Omega_{\rm M}^{\rm flat}$ for the
spiral host-galaxy subset.  Although the consistency of these subset
results is encouraging, the uncertainties are still large enough
(approximately twice the Fit~C uncertainties)
that this test will need to await the host-galaxy
classification of the full set of high-redshift supernovae and a larger
low-redshift supernova sample.

\subsection{Further Cross-Checks}

We have checked several other possible effects that might bias our results,
by fitting different supernova subsets and using alternative analyses:

\vspace{.1in} {\em Sensitivity to Width-Luminosity Correction.}
Although the lightcurve width correction provides some insurance against
supernova evolution biasing our results,
Figure~\ref{stretchhists} shows that almost all
of the Fit~C supernovae at both low- and high-redshift are clustered tightly
around the most-probable value of $s=1$, the standard width for a $B$-band
Leibundgut SN~Ia template lightcurve.
Our results are therefore rather robust with
respect to the actual choice of width-luminosity relation.  We have tested
this sensitivity by re-fitting the supernovae of Fit~C, but with no
width-luminosity correction.  The results (Fit~D), as shown in
Figure~\ref{confmulti}(b), and listed in Table~\ref{fitresults}, are in
extremely close agreement with those of the lightcurve-width-corrected Fit~C.
The statistical uncertainties are also quite close; the
lightcurve-width correction does not significantly improve the
statistical dispersion for the magnitude residuals,
because of the uncertainty in $s$, the measured
lightcurve width.
It is clear that the best-fit cosmology does not depend strongly on the
extra degree of freedom allowed by including the width-luminosity relation
in the fit.

\vspace{.1in} {\em Sensitivity to Non-SN~Ia Contamination.}
We have tested for the possibility of contamination by non-SN~Ia
events masquerading as SNe~Ia in our sample, by performing a fit after
excluding any supernovae with less certain SN~Ia
spectroscopic and photometric identification.
This selection removes
the large statistical outliers from the sample.  In part, this may be
because the host-galaxy contamination that can make it difficult
to identify the supernova spectrum can also increase the odds
of extinction or other systematic uncertainties in photometry.
For this more ``pure'' sample of 43 supernovae, we
find $\Omega_{\rm M}^{\rm flat} = 0.33^{+0.10}_{-0.09}$,
just over half of a standard deviation from Fit~C.

\vspace{.1in} {\em Sensitivity to Galactic Extinction Model.}
Finally, we have tested the effect of the choice
of Galactic extinction model, with a fit using
the model of \cite{bh_red82}, rather than
\cite{schl_red98}.
We find no significant
difference in the best-fit cosmological parameters, although we note
that the extinction near the Galactic pole is somewhat larger in the
\citeauthor{schl_red98} model and this leads to a $\sim$0.03 magnitude
larger average offset
between the low-redshift supernova $B$-band observations and
the high-redshift supernovae $R$-band observations.

\section{Results and Error Budget}

From Table~\ref{fitresults} and Figure~\ref{confmulti}(a),
it is clear that the results of Fits A, B,
and C are quite close to each other, so we can conclude that our
measurement is robust with respect to the choice of these supernova
subsets.
The inclusive Fits~A and B are the fits with the
least subjective selection of the
data.  They already indicate the main cosmological results from this
dataset.  However, to make our results robust with respect to
host-galaxy reddening, we use Fit C as our primary fit in this paper.
For Fit C, we find $\Omega_{\rm M}^{\rm flat}  = 0.28^{+0.09}_{-0.08}$
in a flat universe. Cosmologies with
$\Omega_\Lambda = 0$ are a poor fit to the data, at the 99.8\% confidence
level. The contours of Figure~\ref{40conf} more fully characterize the best-fit
confidence regions.  (The table of this two-dimensional
probability distribution is available at http://www-supernova.lbl.gov/.)

The residual plots of Figure~\ref{hubdiag}(b and c) indicate
that the best-fit $\Omega_{\rm M}^{\rm flat} $ in a flat universe is consistent
across the redshift range of the high-redshift supernovae.  Figure~\ref{hubdiag}(c) shows the residuals normalized by uncertainties;
their scatter can be seen to be typical of a normal-distributed
variable, with the exception of the two outlier supernovae that are removed
from all fits after Fit~A, as discussed above.
Figure~\ref{residhists} compares the magnitude-residual distributions (the
projections of Figure~\ref{hubdiag}b) to the Gaussian distributions expected
given the measurement uncertainties and
an intrinsic dispersion of 0.17 mag.  Both the low- and high-redshift
distributions are consistent with the expected distributions;
the formal calculation of the SN~Ia intrinsic-dispersion
component of the observed magnitude dispersion
($\sigma^2_{\rm intrinsic} = \sigma^2_{\rm observed} -
\sigma^2_{\rm measurement}$) yields
$\sigma_{\rm intrinsic} =0.154 \pm 0.04$ for the low-redshift
distribution and $\sigma_{\rm intrinsic} = 0.157 \pm 0.025$
for the high-redshift distribution.
The $\chi^2$ per degree of freedom for this fit, $\chi^2_\nu = 1.12$,
also indicates that the fit model is a reasonable description of the data.
The narrow intrinsic dispersion---which does not increase at high
redshift---provides additional evidence against an increase in extinction
with redshift.  Even if there is grey dust that dims the supernovae
without reddening them, the dispersion would increase, unless the
dust is distributed very uniformly.

A flat, $\Omega_\Lambda = 0$ cosmology is a quite poor fit to the
data.  The $(\Omega_{\rm M},\Omega_\Lambda) = (1,0)$ line
on Figure~\ref{hubdiag}(b) shows
that 38 out of 42 high-redshift supernovae are fainter than
predicted for this model.  These supernovae would have to be
over 0.4 magnitudes brighter than measured (or the
low-redshift supernovae 0.4 magnitudes fainter) for this model to fit
the data.

The $(\Omega_{\rm M},\Omega_\Lambda) = (0,0)$ upper solid line
on Figure~\ref{hubdiag}(a) shows that the
data are still not a good fit to an ``empty universe,''
with zero mass density and cosmological constant.  The high-redshift
supernovae are as a group fainter than predicted for this cosmology;
in this case, these supernovae would have to be almost 0.15
magnitudes brighter for this empty cosmology
to fit the data, and the discrepancy is even larger for
$\Omega_{\rm M} > 0$.  This is reflected in the high probability (99.8\%)
of $\Omega_\Lambda > 0$.

As discussed in \cite{omol_95}, the slope of the contours in
Figure~\ref{40conf} is a function
of the supernova redshift distribution;
since most of the supernovae reported here are
near $z \sim 0.5$, the confidence region is approximately fit by
$0.8 \, \Omega_{\rm M} - 0.6 \,\Omega_\Lambda \approx -0.2 \pm 0.1$.
(The orthogonal linear combination, which is poorly constrained, is fit by
$0.6 \, \Omega_{\rm M} + 0.8 \,\Omega_\Lambda \approx 1.5 \pm 0.7$.)
In P97, we emphasized
that the well-constrained
linear combination is not parallel to any contour of constant
current-deceleration-parameter, $q_0 = \Omega_{\rm M}/2 - \Omega_\Lambda$;
the accelerating/decelerating universe line of Figure~\ref{ages} shows one
such contour at $q_0 = 0$.  Note that with almost all of the confidence region
above this line, only currently accelerating universes fit the data well.
As more of our
highest redshift supernovae are analyzed, the long dimension of the
confidence region will shorten.

\vspace{.15in}
{\em Error Budget}

Most of the sources of statistical error contribute a statistical
uncertainty to each supernova individually, and are included in
the uncertainties listed in Tables~\ref{scpdata}~and~\ref{hamuydata},
with small correlations between these uncertainties given in
the correlated-error matrices (available at http://www-supernova.lbl.gov).
These supernova-specific statistical uncertainties include the
measurement errors on SN peak magnitude, lightcurve stretch factor,
and absolute photometric calibration.  The two sources of statistical error
that are common to all the supernovae are the intrinsic dispersion
of SN~Ia luminosities after correcting for the width-luminosity
relation, taken as 0.17 mag, and the redshift uncertainty due to
peculiar velocities, which are taken as 300 km s$^{-1}$.  Note that the
statistical error in ${\cal M}_B$ and $\alpha$ are derived quantities
from our four-parameter fits.  By integrating the four-dimensional
probability distributions over these two variables, their
uncertainties are included in the final statistical errors.

All uncertainties that are not included in the statistical error budget
are treated as systematic errors for the purposes of this paper.
In Sections 2 and 4, we have identified and bounded
four potentially significant sources
of systematic uncertainty: (1) the extinction uncertainty for
dust that reddens, bounded
at $<$0.025 magnitudes, the maximal effect of the nine reddest and two
faintest of the high-redshift supernovae; (2)
the difference between the Malmquist bias of the low- and high-redshift
supernovae, bounded at $\le$0.03 magnitudes for low-redshift
supernovae biased intrinsically brighter than high-redshift supernovae, and
$<$0.01 magnitudes for high-redshift supernovae biased brighter than
low-redshift supernovae; (3) the cross-filter $K$-correction uncertainty
of $<$0.02 magnitudes; and (4) the $<$0.01 magnitudes uncertainty 
in $K$ corrections and reddening corrections due to the effect of progenitor
metallicity evolution on the rest-frame $B$-band spectral features.
We take the total identified systematic uncertainty to be the quadrature
sum of the sources: +0.04 magnitudes in the
direction of spuriously fainter high-redshift or brighter low-redshift
supernovae, and $-$0.03 magnitudes in the opposite direction.

Note that we treat the possibility of gravitational lensing due
to small-scale clumping of mass as a separate analysis case, rather
than as a contributing systematic error in our primary analysis; the total
systematic uncertainty applies to this analysis as well.
There are also several more hypothetical sources of systematic
error discussed in Section 4, which are not included in our
calculation of identified systematics.  These include grey dust
[with ${\cal R}_B(z=0.5)\;  > \; 2{\cal R}_B({z=0})$] and any
SN~Ia evolutionary effects that might change the zero point of the
lightcurve width-luminosity relation.  We have presented bounds
and tests for these effects which give preliminary indications
that they are not large sources of uncertainty, but at this time
they remain difficult to quantify, at least partly because the
proposed physical processes and entities that might cause the effects
are not completely defined.

To characterize the effect of the identified systematic uncertainties,
we have refit the supernovae of Fit~C for the
hypothetical case (Fit~J) in which each of the high-redshift supernovae were
discovered to be 0.04 magnitudes brighter than measured, or, equivalently,
the low-redshift supernovae were discovered to be 0.04 magnitudes fainter
than measured.  Figure~\ref{confmulti}(e) and Table~\ref{fitresults} show the
results of this fit.  The best-fit flat-universe $\Omega_{\rm M}^{\rm flat} $
varies from that of Fit~C by 0.05,
less than the statistical error bar.
The probability of $\Omega_\Lambda > 0$ is still over 99\%.
When we fit with the smaller systematic error in the opposite direction
(i.e., high-redshift supernovae discovered to be 0.03 magnitudes fainter
than measured), we find (Fit~I) only a 0.04 shift in
$\Omega_{\rm M}^{\rm flat}$ from Fit~C.

The measurement error of the cosmological parameters has
contributions from both the low- and high-redshift supernova
datasets.  To identify the approximate relative importance of
these two contributory sources, we reanalyzed the Fit~C dataset,
first fitting ${\cal M}_B$ and $\alpha$ to the low-redshift dataset (this is
relatively insensitive to cosmological model), and then fitting
$\Omega_{\rm M}$ and $\Omega_\Lambda$ to the high-redshift dataset.
(This is only an approximation, since it neglects the small
influence of the low-redshift supernovae
on $\Omega_{\rm M}$ and $\Omega_\Lambda$,
and of the high-redshift supernovae on ${\cal M}_B$ and $\alpha$,
in the standard four-parameter fit.)
Figure~\ref{confmulti}(f) shows this $\Omega_{\rm M}$--$\Omega_\Lambda$
fit as a solid contour
(labeled Fit~M), with
the 1-sigma uncertainties on ${\cal M}_B$ and $\alpha$ included with the
systematic uncertainties in the dashed-line confidence contours.
This approach parallels the analyses of
\citeauthor{97ap} (\citeyear{perl97}, \citeyear{perl97a}, \citeyear{97ap}),
and thus also provides a direct comparison with the earlier
results.  We find that the more important
contribution to the uncertainty is currently due to the low-redshift
supernova sample.  If three times as many well-observed low-redshift
supernovae were discovered and included in the analysis, then the
statistical uncertainty from the low-redshift dataset would be smaller
than the other sources of uncertainty.

We summarize the relative statistical and systematic uncertainty
contributions in Table~\ref{scorecard}.

\section{Conclusions and Discussion}

The confidence regions of Figure~\ref{40conf} and
the residual plot of Figure~\ref{hubdiag}(b)
lead to several striking
implications.  First, the data are strongly inconsistent with
the $\Lambda = 0$, flat
universe model (indicated with a circle) that has been the theoretically
favored cosmology.  If the simplest inflationary theories
are correct and the universe is spatially flat, then the supernova
data imply that there is a significant, positive cosmological constant.
Thus, the universe may be flat, {\em or} there may
be little or no cosmological constant, but the data are not consistent with both
possibilities simultaneously.  This is the most unambiguous result of the current dataset.

Second, this dataset directly addresses the age of the universe relative to
the Hubble time, $H_0^{-1}$.  Figure~\ref{ages} shows that the
$\Omega_{\rm M}$--$\Omega_\Lambda$ confidence regions are almost parallel
to contours of constant age.  For any value of the Hubble constant
less than $H_0 = 70$~km~s$^{-1}$~Mpc$^{-1}$,
the implied age of the universe is greater than 13 Gyr, allowing
enough time for the oldest stars in globular clusters to evolve \cite[]{chaboyer98, gratton97}.  Integrating over $\Omega_{\rm M}$ and
$\Omega_\Lambda$, the best fit value of the age in Hubble-time units is
$H_0 t_0 = 0.93 ^{+0.06}_{-0.06}$ or equivalently
$t_0=14.5^{+1.0}_{-1.0} \, (0.63/h)$ Gyr.  The age would be somewhat larger
in a flat universe: $H_0 t_0^{\rm flat} = 0.96 ^{+0.09}_{-0.07}$
or equivalently $t_0^{\rm flat}=14.9^{+1.4}_{-1.1} \, (0.63/h)$ Gyr.

Third, even if the universe is not flat, the confidence regions of
Figure~\ref{40conf}
suggest that the cosmological constant is a
significant constituent of the energy density of the universe.  The
best-fit model (the center of the shaded contours) indicates that the energy
density in the cosmological constant is $\sim$0.5 more
than that in the form of mass energy density.
All of the alternative fits listed in Table~\ref{fitresults}
indicate a positive cosmological constant with confidence levels
of order 99\%, even with the systematic uncertainty included in the
fit or with a clumped-matter metric.

Given the potentially revolutionary nature of this third
conclusion, it is important to reexamine the evidence carefully
to find possible loopholes.
None of the
identified sources of statistical and systematic uncertainty
described in the previous sections could account for the data in a
$\Lambda = 0$ universe.  If the universe does in fact have
zero cosmological constant, then some additional physical effect or
``conspiracy'' of statistical effects must be operative---and must make
the high-redshift supernovae appear almost 0.15 mag ($\sim$15\% in flux)
fainter than the low-redshift supernovae.  At this
stage in the study of SNe~Ia, we consider this unlikely but not
impossible.
For example, as mentioned above, some carefully constructed smooth
distribution of large-grain-size grey dust that evolves similarly for
elliptical and spiral galaxies could evade our current tests.
Also, the full dataset of well-studied SNe~Ia is still
relatively small, particularly at low redshifts, and we would like
to see a more extensive study of SNe~Ia in many different host-galaxy
environments before we consider all plausible loopholes (including
those listed in Table~\ref{scorecard}B) to be closed.

Many of these residual concerns about the measurement can be addressed
with new studies of low-redshift supernovae.
Larger samples of well-studied low-redshift supernovae will
permit detailed analyses of statistically significant SN~Ia
subsamples in differing host environments.  For example, the
width-luminosity relation can be checked and compared for supernovae
in elliptical host galaxies, in the cores of spiral galaxies, and in
the outskirts of spiral galaxies.  This comparison can mimic the
effects of finding high-redshift supernovae with a range of progenitor
ages, metallicities, etc.  So far, the results of such studies with
small statistics has not shown any difference in width-luminosity
relation for this range of environments.
These empirical tests of the SNe~Ia can also be complemented by
better theoretical models.  As the datasets improve,
we can expect to learn more about the physics of
SN~Ia explosions and their dependence on the progenitor environment,
strengthening the confidence in the empirical calibrations.
Finally, new well-controlled, digital searches for SNe~Ia at
low redshift will also be able to further reduce the uncertainties
due to systematics such as Malmquist bias.

\subsection{Comparison with Previous Results}

A comparison with the first supernova measurement of the cosmological
parameters in P97 highlights an important aspect of the current measurement.
As discussed in Section~3, the P97 measurement was
strongly skewed by SN 1994H, one of the two supernovae
that are clear statistical outliers from the current 42-supernova distribution.
If SN 1994H had not been included in the P97 sample, then the
cosmological measurements would have agreed within the 1$\sigma$ error bars with
the current result.  (The small changes in the $K$-corrections discussed
in Section 2 are not a significant factor in arriving at this agreement.)
With the small P97 sample size of seven supernovae (only five of which
were used in the P97 width-corrected analysis), and
somewhat larger measurement uncertainties, it was not possible to distinguish
SN 1994H as the statistical outlier.  It is only with the much larger
current sample size that it is easy to distinguish such outliers on a graph such
as Figure~\ref{hubdiag}(c).

The fact that there are any outliers at all raises one cautionary flag for the
current measurement.  Although neither of the current two outliers is a clearly
aberrant SN~Ia (one has no SN Ia spectral confirmation and the other has
a relatively poor constraint on host-galaxy extinction), we are
watching carefully for such aberrant events in future low- and high-redshift
datasets.
Ideally, the one-parameter width-luminosity relationship
for SNe~Ia would completely account for every single well-studied
SN Ia event.  This is
not a requirement for a robust measurement, but any exceptions that
are discovered would provide an indicator of as-yet undetected
parameters within the main SN Ia distribution.

Our first presentation of the cosmological parameter
measurement \cite[]{perl97a}, based on 40 of the current 42 high-redshift supernovae, found the same basic results as the current
analysis:  A flat universe was shown to require a cosmological
constant, and only a small region of low-mass-density parameter
space, with all the systematic uncertainty included,
could allow for $\Lambda=0$.  (Fit M of Figure~\ref{confmulti}(f) still shows
almost the same confidence region, with the same analysis approach).
The current confidence region of
Figure~\ref{40conf} has changed very little from the corresponding
confidence region of \cite{perl97a}, but since most
of the uncertainties in the low-redshift dataset are now included in
the statistical error, the remaining systematic error is now a
small part of the error budget.

The more recent analysis of 16 high-redshift supernovae by \cite{riess16sne}
also show a very similar $\Omega_{\rm M}$-$\Omega_\Lambda$ confidence region.
The best fits for mass density in a flat-universe  are
$\Omega_{\rm M}^{\rm flat} =
0.28 \pm 0.10$ or $\Omega_{\rm M}^{\rm flat} = 0.16 \pm 0.09$
for the two alternative analyses of their 9 independent,
well-observed, spectroscopically-confirmed supernovae.
The best fits for the age of the universe for these analyses are
$H_0 t_0 = 0.90 ^{+0.07}_{-0.05}$ and $H_0 t_0 = 0.98 ^{+0.07}_{-0.05}$.
To first order, the \citeauthor{riess16sne} result provides an
important independent cross-check for
all three conclusions discussed above, since it was based on a separate
high-redshift supernova search and analysis chain \cite[see][]{schmidtsearch}.
One caveat, however, is that
their $\Omega_{\rm M}$-$\Omega_\Lambda$
confidence-region result cannot be directly compared to ours to check for
independent consistency, because
the low-redshift-supernova datasets are
not independent: a large fraction of these supernovae
with the highest weight in both analyses are from the
Cal\'{a}n/Tololo Supernova Survey (which provided many
well-measured supernovae that were far enough into the Hubble flow so
that their peculiar velocities added negligible redshift-uncertainty).
Moreover, two of the 16 high-redshift
supernovae included in the \citeauthor{riess16sne}
confidence-region analyses were from our sample of 42
Supernova Cosmology Project supernovae; Riess et al. included them
with an alternative analysis technique applied to a subset of our
photometry results. (In particular, their result uses
the highest-redshift supernova from our 42-supernova sample, which has
strong weight in our analysis due to the excellent Hubble
Space Telescope photometry.)
Finally, although the analysis techniques are mostly independent,
the $K$ corrections are based on the same \cite{nuge_kcorr98} approach
discussed above.

\subsection{Comparison with Complementary Constraints \\
on $\Omega_{\rm M}$ and $\Omega_\Lambda$}

Significant progress is being made in the measurement of the cosmological
parameters using complementary techniques that are sensitive to different
linear combinations of $\Omega_{\rm M}$ and $\Omega_\Lambda$, and have
different potential systematics or model dependencies.
Dynamical methods, for example, are particularly sensitive to
$\Omega_{\rm M}$, since $\Omega_\Lambda$ affects dynamics only weakly.
Since there is evidence that dynamical estimates of $\Omega_{\rm M}$
depend on scale, the most appropriate measures to compare with our
result are those obtained on large scales. From the abundance---indeed
the mere existence---of rich clusters at high redshift,
\cite{bahfan98} find $\Omega_{\rm M} = 0.2^{+0.3}_{-0.1}$ (95\%
confidence).  The CNOC collaboration \cite[]{carlberg96, carlberg98}
apply evolution-corrected mass-to-light ratios
determined from virial mass estimates of X-ray clusters to the
luminosity density of the universe and find $\Omega_{\rm M}= 0.17 \pm
0.07$ for $\Omega_\Lambda$ = 0 ($\sim$90\% confidence), with small
changes in these results for different values of $\Omega_\Lambda$
\cite[cf.][]{carlberg97}.  Detailed studies of the peculiar velocities of
galaxies (e.g., \citeauthor{willick97} \citeyear{willick97};
\citeauthor{willick98} \citeyear{willick98};
\citeauthor{riess_beta97} \citeyear{riess_beta97}; but see \citeauthor{sigad98} \citeyear{sigad98})
are now giving estimates of $\beta = \Omega_{\rm
M}^{0.6}/b_{\rm IRAS} \approx 0.45 \pm 0.11$ (95\% confidence)$^1$, where
$b$ is the ratio of density contrast in galaxies compared to that in
all matter.  Under the simplest assumption of no large-scale biasing
for IRAS galaxies, $b = 1$, these results give $\Omega_{\rm M}\approx 0.26
\pm 0.11$ (95\% confidence), in agreement with the other dynamical
estimates---and with our supernova results for a flat cosmology.

\footnotetext[1]{This is an error-weighted
mean of \cite{willick97} and \cite{riess_beta97}, with optical
results converted to equivalent IRAS results
using $b_{\rm Opt}/b_{\rm IRAS} =
1.20 \pm 0.05$ from \cite{oliver96}.}

A form of the angular-size distance cosmological test has been
developed in a series of papers
\cite[cf.][and references therein]{guerra98}
and implemented for a sample of fourteen radio galaxies by
\cite{daly98}. The method uses the mean observed separation of the
radio lobes compared to a canonical maximum lobe size---calculated
from the inferred magnetic field strength, lobe propagation velocity,
and lobe width---as a calibrated standard ruler.  The confidence
region in the $\Omega_{\rm M}$--$\Omega_\Lambda$ plane shown in
\cite{daly98} is in broad agreement with the SN~Ia results we report;
they find $\Omega_{\rm M} = 0.2^{+0.3}_{-0.2}$ (68\% confidence) for a
flat cosmology.

QSO gravitational lensing statistics are dependent on both volume and
relative distances, and thus are more sensitive to $\Omega_\Lambda$.
Using gravitational lensing statistics, \cite{kochanek96} finds
$\Omega_\Lambda < 0.66$ (at 95\% confidence for
$\Omega_{\rm M}+\Omega_\Lambda$ = 1), and $\Omega_{\rm M} > 0.15$.
\cite{falco97} obtained further information on the
redshift distribution of radio sources which allows calculation of the
absolute lensing probability for both optical and radio lenses. Formally
their 90\% confidence levels in the $\Omega_{\rm M}$--$\Omega_\Lambda$
plane have no overlap with those we report here. However, as \cite{falco97}
discuss, these results do depend on the choice of galaxy sub-type
luminosity functions in the lens models.
\cite{chiba98} emphasized this point, reporting an analysis with
E/S0 luminosity functions that yielded a best-fit mass
density in a flat cosmology of
$\Omega_{\rm M}^{\rm flat} = 0.3 ^{+0.2}_{-0.1}$, in agreement
with our SN~Ia results.

Several papers have emphasized that upcoming balloon and satellite
studies of the Cosmic Background Radiation (CBR) should provide a good
measurement of the sum of the energy densities, $\Omega_{\rm M}$ +
$\Omega_\Lambda$, and thus provide almost orthogonal information to the
supernova measurements \cite[]{white98, tegmark98}.  In
particular, the position of the first acoustic peak in the CBR
power spectrum is sensitive to this combination of the cosmological
parameters.  The current results, while not conclusive, are already
somewhat inconsistent with over-closed ($\Omega_{\rm M}+
\Omega_\Lambda >\!> 1$) cosmologies and ``near-empty''  ($\Omega_{\rm M}+
\Omega_\Lambda \lesssim 0.4$) cosmologies, and may exclude the upper right
and lower left
regions of Figure~\ref{40conf}
\cite[see, e.g.,][]{lineweaver98,efstathiouprep}.

\subsection{Cosmological Implications}

If, in fact, the universe has a dominant energy contribution from a
cosmological constant, there are two coincidences that must be addressed
in future cosmological theories.  First, a cosmological constant
in the range shown in Figure~\ref{40conf} corresponds to a very
small energy density relative to the vacuum-energy-density scale of
particle-physics energy zero-points
\cite[see][for a discussion of this point]{carrollpressturner}.
Previously, this had been seen as an argument for a zero cosmological
constant, since presumably some symmetry of the particle-physics model
is causing cancelations of this vacuum energy density.
Now, it would be necessary to explain how this value comes to be so small,
yet non-zero.

Second, there is the coincidence that the cosmological constant value
is comparable to the current mass-energy density.  As the universe
expands, the matter energy density falls as the third power of the scale,
while the cosmological constant remains unchanged.
One therefore would require initial conditions in which the ratio of densities
is a special, infinitesimal value of order $10^{-100}$ in order for
the two densities to coincide today.
(The cross-over between mass-dominated and $\Lambda$-dominated
energy density occurred at $z \approx 0.37$, for a flat
$\Omega_{\rm M} \approx 0.28$ universe, whereas the cross-over
between deceleration and acceleration
occurred when $(1+z)^3 \Omega_{\rm M}/2 = \Omega_\Lambda$, that is at
$z \approx 0.73$.  This was approximately
when SN 1997G exploded, over 6 billion years ago.)

It has been suggested that these cosmological coincidences could be
explained if the magnitude-redshift relation we find
for SNe~Ia is due not to a cosmological constant, but rather
to a different, previously unknown physical
entity that contributes to the universe's total energy
density \cite[see, e.g.,][]{stein96,turnerwhite97,caldwell98}.
Such an entity can lead to a different
expansion history than the cosmological constant does, because it can
have a different relation (``equation of state'') between its density
$\rho$ and pressure $p$ than that of the cosmological constant,
$p_\Lambda/\rho_\Lambda = -1$.
We can obtain constraints on this equation-of-state
ratio, $w \equiv p/\rho$, and check for consistency with
alternative theories
(including the cosmological constant with $w=-1$) by
fitting the alternative expansion histories to data;
\cite{white98} has
discussed such constraints
from earlier supernovae and
CBR results. In Figure~\ref{wconf},
we update these constraints for our current supernova
dataset, for the simplest case of a flat universe
and an equation of state that does not vary in time
\cite[cf.][for comparison with their high-redshift
supernova dataset,
and Aldering et al. 1998 for time-varying equations of
state fit to our dataset]{garnavichwfit}.
In this simple case, a cosmological-constant
equation of state can fit our data if the mass density
is in the range $0.2 \lesssim \Omega_{\rm M} \lesssim 0.4$.
However, all the cosmological models shown in Figure~\ref{wconf}
still require that
the initial conditions for the new energy density be tuned with
extreme precision to reach their current-day values.
\cite{zlatev98} have shown that some
time-varying-$w$ theories naturally channel the new energy density term
to ``track'' the
matter term, as the universe expands, leading---without
coincidences---to values of an
effective vacuum energy density today that are comparable
to the mass energy density.  These models require $w \gtrsim -0.8$
at all times up to the present,
for $\Omega_{\rm M} \ge 0.2$.
The supernova dataset
presented here and future complementary datasets
will allow us to explore these possibilities.

\vspace{0.8in}

\acknowledgements

The observations described in this paper
were primarily obtained as visiting/guest astronomers at
the Cerro Tololo
Inter-American Observatory 4-meter telescope, operated by the
National Optical Astronomy Observatory under contract to the National
Science Foundation; the Keck I and II 10-m telescopes of  the California
Association for Research in Astronomy;
the Wisconsin-Indiana-Yale-NOAO (WIYN) telescope;
the European Southern Observatory 3.6-meter telescope;
the Isaac
Newton and William Herschel Telescopes, operated by the Royal Greenwich
Observatory at the Spanish Observatorio del Roque de los Muchachos of
the Instituto de Astrofisica de Canarias;
the Hubble Space Telescope, and the Nordic Optical 2.5-meter
telescope.  We thank the dedicated staff of these observatories for
their excellent assistance in pursuit of this project.
In particular, Dianne Harmer, Paul
Smith and Daryl Willmarth were extraordinarily helpful as the WIYN queue
observers.   We thank Gary Bernstein and Tony Tyson for developing and
supporting the Big Throughput Camera at the CTIO 4-meter; this wide-field
camera was important in the discovery of many of the high-redshift supernovae.
David Schlegel, Doug Finkbeiner, and Marc Davis provided early
access to, and helpful discussions concerning, their models of
Galactic extinction.  Megan Donahue contributed serendipitous
HST observations of SN 1996cl.  We thank Daniel Holz and Peter
H\"{o}flich for helpful discussions.
The larger computations described in this paper were performed at the
U.~S. Department of Energy's National Energy Research Science Computing
Center (NERSC).
This work was supported in part by the
Physics Division, E.~O. Lawrence Berkeley National Laboratory of the
U.~S. Department of Energy under Contract No. DE-AC03-76SF000098, and
by the National Science Foundation's Center for Particle Astrophysics,
University of California, Berkeley under grant No. ADT-88909616.
A.~V.~F. acknowledges the support of NSF grant No. AST-9417213 and
A.~G. acknowledges the support of the Swedish Natural Science Research
Council.  The France-Berkeley Fund and the Stockholm-Berkeley Fund
provided additional collaboration support.

\vspace{0.5in}

\appendix

\section{Extinction Correction using a Bayesian Prior}

 Bayes Theorem provides a means of estimating the {\it a posteriori}
probability distribution, $P(A \vert A_m)$, of a variable $A$
given a measurement of its value, $A_m$, along with {\it a priori}
information, $P(A)$, about what values are likely:

\begin{equation}
P(A \vert A_m) = {{P(A_m \vert A) P(A)}\over{\int{P(A_m \vert A) P(A) dA}}}
\end{equation}

 In practice $P(A)$ often is not well known, but must be estimated from
sketchy, and possibly biased, data. For our purposes here we wish to
distinguish between the true probability distribution, $P(A)$,  and its
estimated or assumed distribution, often called the Bayesian
prior, which we denote as ${\cal P}(A)$.
 Riess, Press, \& Kirshner (1996; RPK) present a Bayesian method of
correcting SNe~Ia for host galaxy extinction. For ${\cal P}(A)$ they assume
a one-sided Gaussian function of extinction, ${\cal G}(A)$, with dispersion
$\sigma_{\cal G} = 1$ magnitude:

\begin{equation}
{\cal P}(A) = {\cal G}(A) \equiv \left\{ \begin{array}{ll}
{\displaystyle
\sqrt{{{2}\over{\pi \sigma_{\cal G}^2}}}
{\displaystyle e^{\scriptstyle {-A^2}/2\sigma_{\cal G}^2}}}
& \mbox{for $A \geq 0$} \\
\\
 0  & \mbox{for $A < 0$}
\end{array}
\right.
\label{onesided}
\end{equation}

\noindent
which reflects the fact that dust can only redden and dim the light
from a supernova.  The probability distribution of the measured
extinction, $A_m$, is an ordinary Gaussian with dispersion $\sigma_m$,
i.e., the measurement uncertainty.  RPK
choose the most probable value of $P(A \vert A_m)$ as their best
estimate of the extinction for each supernova:

\begin{equation}
{\hat{A}_{\cal G}} =
{\rm mode}(P(A \vert A_m)) = \left\{ \begin{array}{ll}
{\displaystyle A_m {\sigma_{\cal G}^2}
\over{\displaystyle \sigma_{\cal G}^2 + \sigma_m^2}}
& \mbox{for $A_m > 0$} \\
\\
 0  & \mbox{for $A_m \leq 0$}
\end{array}
\right.
\end{equation}

Although this method provides the best estimate of the extinction correction
for an individual supernova provided ${\cal P}(A) = P(A)$, once measurement
uncertainties are considered its application to an ensemble of SNe~Ia
can result in a biased estimate of the ensemble average extinction whether or
not ${\cal P}(A) = P(A)$. An extreme case which illustrates this point is
where the true extinction is
zero for all supernovae, i.e., P(A) is a delta function at zero.
In this case, a measured value of $E(B\!-\!V) < 0 $ (too
blue) results in an extinction estimate of ${\hat{A}_{\cal G}} = 0$, while a
measured value with $E(B\!-\!V) \geq 0 $ results in an extinction
estimate ${\hat{A}_{\cal G}} > 0$. The ensemble mean of these extinction
estimates will be

\begin{equation}
\langle {\hat{A}_{\cal G}}\rangle = {{\sigma_m}\over{\sqrt{2\pi}}} ({{\sigma_{\cal G}^2}\over{\sigma_{\cal G}^2 + \sigma_m^2}}),
\end{equation}

\noindent
rather than 0 as it should be. (This result
is changed only slightly if the smaller uncertainties assigned to the least
extincted SNe~Ia are incorporated into a weighted average.)

The amount of this bias is
dependent on the size of the extinction-measurement uncertainties,
$\sigma_m = {\cal R}_B \sigma_{E(B\!-\!V)}$.
For our sample of high-redshift supernovae, typical values of
this uncertainty are $\sigma_m \sim 0.5$, while for the low-redshift
supernovae, $\sigma_m \sim 0.07$.
Thus, if the
true extinction distribution is a delta-function at $A = 0$,
while the one-sided prior, ${\cal G}(A)$, of Equation~\ref{onesided} is used,
the bias in $\langle {\hat{A}_{\cal G}}\rangle$ is about 0.13
mag in the sense that the high-redshift supernovae would be overcorrected
for extinction.
Clearly, the
exact amount of bias depends on the details of the dataset (e.g., color
uncertainty, relative weighting). the true distribution $P(A)$, and
the choice of prior ${\cal P}(A)$.
This is a worst-case estimate, since we believe that
the true extinction distribution is more likely to have some tail
of events with extinction.
Indeed, numerical calculations using a one-sided Gaussian for the true
distribution,
$P(A)$, show that
the amount of bias decreases as the Gaussian width increases away from
a delta function, crosses zero when $P(A)$ is still {\it much narrower} than ${\cal P}(A)$, and then increases with opposite sign.
One might use the mean of $P(A \vert A_m)$ instead of the mode
in equation A-3, since the bias then vanishes if ${\cal P}(A) = P(A)$,
however this mean-calculated bias is even more sensitive to
${\cal P}(A) \neq P(A)$ than the mode-calculated bias.

We have only used conservative priors (which are somewhat broader than
the true distribution, as discussed in Section 4.1), however it is
instructive to consider the bias that results for a less
conservative choice of prior.  For example,
an extinction distribution with only half of the supernovae distributed
in a one-sided Gaussian and half in a delta function at zero extinction
is closer to the simulations given by \cite{hatano97}.
The presence of the delta-function
component in this less conservative prior assigns zero extinction to the
vast majority of supernovae, and thus
cannot produce a bias even with different uncertainties at low and high
redshift.  This will lower the overall bias, but it will also
assign zero extinction to many more supernovae than assumed in the prior,
in typical cases in which the measurement uncertainty is not significantly
smaller than the true extinction distribution.  A restrictive prior,
i.e. one which is actually narrower than the true distribution, can even
lead to a bias in the opposite direction from a conservative prior.

It is clear from Bayes Theorem itself that the correct procedure for
determining the maximum-likelihood extinction, $\tilde{A}$, of an
ensemble of supernovae is to first calculate the {\it a posteriori} probability
distribution for the ensemble:

\begin{equation}
 P(A \vert \{A_{m_i}\}) =
{{P(A) \sum P(A_{m_i} \vert A )}\over{\int{P(A) \sum P(A_{m_i} \vert A) dA}}}
\end{equation}

\noindent
and then take the most probable value of $P(A \vert \{A_{m_i}\})$ for
$\tilde{A}$. For the above example of no reddening, this returns the
correct value of $\tilde{A} = 0$.

In fitting the cosmological parameters generally one is not quite as
interested in the ensemble extinction as in the combined impact of
individual extinctions. In this case $P(A \vert \{A_{m_i}\})$ must be
combined with other sources of uncertainty for each supernova in a
maximum--likelihood fit, or the use of a Bayesian prior must be
abandoned. In the former case a $\chi^2$ fit is no longer appropriate
since the individual $P(A \vert \{A_{m_i}\})$'s are strongly
non-Gaussian.
Use of a Gaussian uncertainty for ${\hat{A}_{\cal G}}$ based on the
second--moment of $P(A \vert \{A_{m_i}\})$ may introduce additional biases.

\clearpage


\clearpage

\def\thesection{\@}


\begin{deluxetable}{lcccccccccl}
\scriptsize
\tablenum{1}

\tablecaption{SCP SNe~Ia Data \label{scpdata}}
\tablehead{
\colhead{SN} &
\colhead{$z$} &
\colhead{$\sigma_z$} &
\colhead{$m_{X}^{\rm peak}$} &
\colhead{$\sigma_{X}^{\rm peak}$} &
\colhead{$A_X$} &
\colhead{$K_{BX}$} &
\colhead{$m_{B}^{\rm peak}$} &
\colhead{$m_{B}^{\rm effective}$} &
\colhead{$\sigma_{m_{B}^{\rm effective}}$} &
\colhead{Notes} \\
\colhead{$(1)$} &
\colhead{$(2)$} &
\colhead{$(3)$} &
\colhead{$(4)$} &
\colhead{$(5)$} &
\colhead{$(6)$} &
\colhead{$(7)$} &
\colhead{$(8)$} &
\colhead{$(9)$} &
\colhead{$(10)$} &
\colhead{$(11)$}
}
\startdata
1992bi & 0.458 & 0.001 & 22.12 & 0.10 & 0.03 & $-$0.72 & 22.81 & 23.11 & 0.46 & E--H    \\[-0.0ex]
1994F  & 0.354 & 0.001 & 22.08 & 0.10 & 0.11 & $-$0.58 & 22.55 & 22.38 & 0.33 & E--H    \\[-0.0ex]
1994G  & 0.425 & 0.001 & 21.52 & 0.21 & 0.03 & $-$0.68 & 22.17 & 22.13 & 0.49 &         \\[-0.0ex]
1994H  & 0.374 & 0.001 & 21.28 & 0.06 & 0.10 & $-$0.61 & 21.79 & 21.72 & 0.22 & B--L    \\[-0.0ex]
1994al & 0.420 & 0.001 & 22.37 & 0.06 & 0.42 & $-$0.68 & 22.63 & 22.55 & 0.25 & E--H    \\[-0.0ex]
1994am & 0.372 & 0.001 & 21.82 & 0.07 & 0.10 & $-$0.61 & 22.32 & 22.26 & 0.20 & E--H    \\[-0.0ex]
1994an & 0.378 & 0.001 & 22.14 & 0.08 & 0.21 & $-$0.62 & 22.55 & 22.58 & 0.37 & E--H    \\[-0.0ex]
1995aq & 0.453 & 0.001 & 22.60 & 0.07 & 0.07 & $-$0.71 & 23.24 & 23.17 & 0.25 &         \\[-0.0ex]
1995ar & 0.465 & 0.005 & 22.71 & 0.04 & 0.07 & $-$0.71 & 23.35 & 23.33 & 0.30 & H       \\[-0.0ex]
1995as & 0.498 & 0.001 & 23.02 & 0.07 & 0.07 & $-$0.71 & 23.66 & 23.71 & 0.25 & H       \\[-0.0ex]
1995at & 0.655 & 0.001 & 22.62 & 0.03 & 0.07 & $-$0.66 & 23.21 & 23.27 & 0.21 & H       \\[-0.0ex]
1995aw & 0.400 & 0.030 & 21.75 & 0.03 & 0.12 & $-$0.65 & 22.27 & 22.36 & 0.19 &         \\[-0.0ex]
1995ax & 0.615 & 0.001 & 22.53 & 0.07 & 0.11 & $-$0.67 & 23.10 & 23.19 & 0.25 &         \\[-0.0ex]
1995ay & 0.480 & 0.001 & 22.64 & 0.04 & 0.35 & $-$0.72 & 23.00 & 22.96 & 0.24 &         \\[-0.0ex]
1995az & 0.450 & 0.001 & 22.44 & 0.07 & 0.61 & $-$0.71 & 22.53 & 22.51 & 0.23 &         \\[-0.0ex]
1995ba & 0.388 & 0.001 & 22.08 & 0.04 & 0.06 & $-$0.63 & 22.66 & 22.65 & 0.20 &         \\[-0.0ex]
1996cf & 0.570 & 0.010 & 22.70 & 0.03 & 0.13 & $-$0.68 & 23.25 & 23.27 & 0.22 &         \\[-0.0ex]
1996cg & 0.490 & 0.010 & 22.46 & 0.03 & 0.11 & $-$0.72 & 23.06 & 23.10 & 0.20 & C,D,G--L\\[-0.0ex]
1996ci & 0.495 & 0.001 & 22.19 & 0.03 & 0.09 & $-$0.71 & 22.82 & 22.83 & 0.19 &         \\[-0.0ex]
1996ck & 0.656 & 0.001 & 23.08 & 0.07 & 0.13 & $-$0.66 & 23.62 & 23.57 & 0.28 &         \\[-0.0ex]
1996cl & 0.828 & 0.001 & 23.53 & 0.10 & 0.18 & $-$1.22 & 24.58 & 24.65 & 0.54 &         \\[-0.0ex]
1996cm & 0.450 & 0.010 & 22.66 & 0.07 & 0.15 & $-$0.71 & 23.22 & 23.17 & 0.23 &         \\[-0.0ex]
1996cn & 0.430 & 0.010 & 22.58 & 0.03 & 0.08 & $-$0.69 & 23.19 & 23.13 & 0.22 & C,D,G--L\\[-0.0ex]
1997F  & 0.580 & 0.001 & 22.90 & 0.06 & 0.13 & $-$0.68 & 23.45 & 23.46 & 0.23 & H       \\[-0.0ex]
1997G  & 0.763 & 0.001 & 23.56 & 0.41 & 0.20 & $-$1.13 & 24.49 & 24.47 & 0.53 &         \\[-0.0ex]
1997H  & 0.526 & 0.001 & 22.68 & 0.05 & 0.16 & $-$0.70 & 23.21 & 23.15 & 0.20 & H       \\[-0.0ex]
1997I  & 0.172 & 0.001 & 20.04 & 0.02 & 0.16 & $-$0.33 & 20.20 & 20.17 & 0.18 &         \\[-0.0ex]
1997J  & 0.619 & 0.001 & 23.25 & 0.08 & 0.13 & $-$0.67 & 23.80 & 23.80 & 0.28 &         \\[-0.0ex]
1997K  & 0.592 & 0.001 & 23.73 & 0.10 & 0.07 & $-$0.67 & 24.33 & 24.42 & 0.37 & H       \\[-0.0ex]
1997L  & 0.550 & 0.010 & 22.93 & 0.05 & 0.08 & $-$0.69 & 23.53 & 23.51 & 0.25 &         \\[-0.0ex]
1997N  & 0.180 & 0.001 & 20.19 & 0.01 & 0.10 & $-$0.34 & 20.42 & 20.43 & 0.17 & H       \\[-0.0ex]
1997O  & 0.374 & 0.001 & 22.97 & 0.07 & 0.09 & $-$0.61 & 23.50 & 23.52 & 0.24 & B--L    \\[-0.0ex]
1997P  & 0.472 & 0.001 & 22.52 & 0.04 & 0.10 & $-$0.72 & 23.14 & 23.11 & 0.19 &         \\[-0.0ex]
1997Q  & 0.430 & 0.010 & 22.01 & 0.03 & 0.09 & $-$0.69 & 22.60 & 22.57 & 0.18 &         \\[-0.0ex]
1997R  & 0.657 & 0.001 & 23.28 & 0.05 & 0.11 & $-$0.66 & 23.83 & 23.83 & 0.23 &         \\[-0.0ex]
1997S  & 0.612 & 0.001 & 23.03 & 0.05 & 0.11 & $-$0.67 & 23.59 & 23.69 & 0.21 &         \\[-0.0ex]
1997ac & 0.320 & 0.010 & 21.38 & 0.03 & 0.09 & $-$0.55 & 21.83 & 21.86 & 0.18 &         \\[-0.0ex]
1997af & 0.579 & 0.001 & 22.96 & 0.07 & 0.09 & $-$0.68 & 23.54 & 23.48 & 0.22 &         \\[-0.0ex]
1997ai & 0.450 & 0.010 & 22.25 & 0.05 & 0.14 & $-$0.71 & 22.81 & 22.83 & 0.30 & H       \\[-0.0ex]
1997aj & 0.581 & 0.001 & 22.55 & 0.06 & 0.11 & $-$0.68 & 23.12 & 23.09 & 0.22 &         \\[-0.0ex]
1997am & 0.416 & 0.001 & 21.97 & 0.03 & 0.11 & $-$0.67 & 22.52 & 22.57 & 0.20 &         \\[-0.0ex]
1997ap & 0.830 & 0.010 & 23.20 & 0.07 & 0.13 & $-$1.23 & 24.30 & 24.32 & 0.22 & H       \\[-0.0ex]
\enddata
\tablenotetext{}{
\vspace{-.25in} 
\,\newline
Col  1. IAU Name assigned to SCP supernova.\newline
Col  2. Geocentric redshift of supernova or host galaxy.\newline
Col  3. Redshift uncertainty.\newline
Col  4. Peak magnitude from lightcurve fit in observed band corresponding to restframe $B$-band
 (i.e. $m_{X}^{\rm peak}\equiv m_{R}^{\rm peak}$ or $m_{I}^{\rm peak}$).\newline
Col  5. Uncertainty in fit peak magnitude.\newline
Col  6. Galactic extinction in observed band
corresponding to restframe $B$-band (i.e., $A_X \equiv A_R$ or $A_I$);
an uncertainty of 10\% is assumed.\newline
Col  7. Representative $K$-correction (at peak) from observed band
to $B$-band (i.e., $K_{BX} \equiv K_{BR}$ or $K_{BI}$);
an uncertainty of 2\% is assumed.\newline
Col  8. $B$-band peak magnitude.\newline
Col  9. Stretch-luminosity corrected effective $B$-band peak magnitude:
$m_B^{\rm effective} \equiv m_{X}^{\rm peak} + \alpha(s-1) - K_{BX} - A_X$.\newline
Col 10. Total uncertainty in corrected $B$-band peak magnitude. This includes uncertainties due to
        the intrinsic luminosity dispersion of SNe~Ia of 0.17 mag, 10\% of the Galactic extinction
        correction, 0.01 mag for $K$-corrections, 300 km s$^{-1}$ to account for peculiar velocities,
        in addition to propagated uncertainties from the lightcurve fits.\newline
Col 11. Fits from which given supernova was excluded.\newline
}
\end{deluxetable}

\begin{deluxetable}{lcccccccccl}
\scriptsize
\tablenum{2}
\tablecaption{Cal\'{a}n Tololo SNe~Ia Data \label{hamuydata}}
\tablehead{
\colhead{SN} &
\colhead{$z$} &
\colhead{$\sigma_z$} &
\colhead{$m_{obs}^{\rm peak}$} &
\colhead{$\sigma_{obs}^{\rm peak}$} &
\colhead{$A_B$} &
\colhead{$K_{BB}$} &
\colhead{$m_{B}^{\rm peak}$} &
\colhead{$m_{B}^{\rm corr}$} &
\colhead{$\sigma_{m_{B}^{\rm corr}}$ } &
\colhead{Notes} \\
\colhead{$(1)$} &
\colhead{$(2)$} &
\colhead{$(3)$} &
\colhead{$(4)$} &
\colhead{$(5)$} &
\colhead{$(6)$} &
\colhead{$(7)$} &
\colhead{$(8)$} &
\colhead{$(9)$} &
\colhead{$(10)$} &
\colhead{$(11)$}
}
\startdata
1990O  & 0.030 & 0.002 & 16.62 & 0.03 & 0.39 &$-$0.00 & 16.23 & 16.26 & 0.20 & \\
1990af & 0.050 & 0.002 & 17.92 & 0.01 & 0.16 &$+$0.01 & 17.75 & 17.63 & 0.18 & \\
1992P  & 0.026 & 0.002 & 16.13 & 0.03 & 0.12 &$-$0.01 & 16.02 & 16.08 & 0.24 & \\
1992ae & 0.075 & 0.002 & 18.61 & 0.12 & 0.15 &$+$0.03 & 18.43 & 18.43 & 0.20 & \\
1992ag & 0.026 & 0.002 & 16.59 & 0.04 & 0.38 &$-$0.01 & 16.22 & 16.28 & 0.20 & \\
1992al & 0.014 & 0.002 & 14.60 & 0.01 & 0.13 &$-$0.01 & 14.48 & 14.47 & 0.23 & \\
1992aq & 0.101 & 0.002 & 19.29 & 0.12 & 0.05 &$+$0.05 & 19.19 & 19.16 & 0.23 & \\
1992bc & 0.020 & 0.002 & 15.20 & 0.01 & 0.07 &$-$0.01 & 15.13 & 15.18 & 0.20 & \\
1992bg & 0.036 & 0.002 & 17.41 & 0.07 & 0.77 &$+$0.00 & 16.63 & 16.66 & 0.21 & \\
1992bh & 0.045 & 0.002 & 17.67 & 0.04 & 0.10 &$+$0.01 & 17.56 & 17.61 & 0.19 & \\
1992bl & 0.043 & 0.002 & 17.31 & 0.07 & 0.04 &$+$0.01 & 17.26 & 17.19 & 0.18 & \\
1992bo & 0.018 & 0.002 & 15.85 & 0.02 & 0.11 &$-$0.01 & 15.75 & 15.61 & 0.21 & B--L\\
1992bp & 0.079 & 0.002 & 18.55 & 0.02 & 0.21 &$+$0.04 & 18.30 & 18.27 & 0.18 & \\
1992br & 0.088 & 0.002 & 19.71 & 0.07 & 0.12 &$+$0.04 & 19.54 & 19.28 & 0.18 & B--L\\
1992bs & 0.063 & 0.002 & 18.36 & 0.05 & 0.09 &$+$0.03 & 18.24 & 18.24 & 0.18 & \\
1993B  & 0.071 & 0.002 & 18.68 & 0.08 & 0.31 &$+$0.03 & 18.34 & 18.33 & 0.20 & \\
1993O  & 0.052 & 0.002 & 17.83 & 0.01 & 0.25 &$+$0.01 & 17.58 & 17.54 & 0.18 & \\
1993ag & 0.050 & 0.002 & 18.29 & 0.02 & 0.56 &$+$0.01 & 17.71 & 17.69 & 0.20 & \\
\enddata
\tablenotetext{}{
\,\newline
Col  1. IAU Name assigned to Cal\'{a}n Tololo supernova.\newline
Col  2. Redshift of supernova or host galaxy in Local Group restframe.\newline
Col  3. Redshift uncertainty.\newline
Col  4. Peak magnitude from lightcurve fit, in observed $B$-band.
Note that the template lightcurve used in the fit is not identical to the 
template lightcurve used by Hamuy et al. so the best-fit peak magnitude may
differ slightly.\newline
Col  5. Uncertainty in fit peak magnitude.\newline
Col  6. Galactic extinction in observed $B$-band; an uncertainty of 10\% is assumed.\newline
Col  7. Representative $K$-correction from observed $B$-band to restframe $B$-band;
an uncertainty of 2\% is assumed.\newline
Col  8. $B$-band peak magnitude.\newline
Col  9. Stretch--luminosity corrected $B$-band peak magnitude.\newline
Col 10. Total uncertainty in corrected $B$-band peak magnitude. This includes uncertainties due to
        the intrinsic luminosity dispersion of SNe~Ia of 0.17 mag, 10\% of the Galactic extinction
        correction, 0.01 mag for $K$--corrections, 300 km s$^{-1}$ to account for peculiar velocities,
        in addition to propagated uncertainties from the lightcurve fits.\newline
Col 11. Fits from which given supernova was excluded.\newline
}
\end{deluxetable}

\begin{deluxetable}{c c c c c c c l}
\textwidth=9in \textheight=8in \hoffset=0in \voffset=-.25in

\scriptsize
\tablecaption{Fit Results \label{fitresults}}
\tablenum{3}
\tablehead{Fit & N & $\chi^2$ & DOF &
           $\Omega_{\rm M}^{\rm flat}$ & $P(\Omega_\Lambda > 0)$ & Best Fit & Fit Description \\
            & & & & & & $\Omega_{\rm M},\Omega_{\Lambda}$  & }

\startdata
\multicolumn{8}{l}{\em Inclusive Fits}\\
A & 60 & 98 & 56 & 0.29$^{+0.09}_{-0.08}$ & 0.9984 & 0.83,1.42 & All SNe \\[+0.5ex]
B & 56 & 60 & 52 & 0.26$^{+0.09}_{-0.08}$ & 0.9992 & 0.85,1.54 & Fit A, but excluding 2 residual \\[-0.3ex]
  &    &    &    &                        &        &           & outliers and 2 stretch outliers\\
\\
\tableline
\multicolumn{8}{l}{\em Primary Fit}\\
{\bf C} & {\bf 54} & {\bf 56} & {\bf 50} & ${\bf 0.28^{+0.09}_{-0.08}}$ & {\bf 0.9979}& {\bf 0.73,1.32} & {\bf Fit B, but also excluding}\\[-0.3ex]
        &          &          &          &                              &             &                 & {\bf 2 likely reddened}\\
\tableline
\\
\multicolumn{8}{l}{\em Comparison Analysis Techniques}\\
D & 54 & 53 & 51 & 0.25$^{+0.10}_{-0.09}$ & 0.9972 & 0.76,1.48 & No stretch correction\tablenotemark{a}\\[+0.5ex]
E & 53 & 62 & 49 & 0.29$^{+0.12}_{-0.10}$ & 0.9894 & 0.35,0.76 & Bayesian one-sided extinction corrected\tablenotemark{b}\\
\\
\multicolumn{8}{l}{\em Effect of Reddest SNe}\\
F & 51 & 59 & 47 & 0.26$^{+0.09}_{-0.08}$ & 0.9991 & 0.85,1.54 & Fit B SNe with colors measured\\[+0.5ex]
G & 49 & 56 & 45 & 0.28$^{+0.09}_{-0.08}$ & 0.9974 & 0.73,1.32 & Fit C SNe with colors measured\\[+0.5ex]
H & 40 & 33 & 36 & 0.31$^{+0.11}_{-0.09}$ & 0.9857 & 0.16,0.50 & Fit G, but excluding 7 next reddest\\
 &    &    &    &                        &        &           & and 2 next faintest high-redshift SNe\\
\\
\multicolumn{8}{l}{\em Systematic Uncertainty Limits}\\
I & 54 & 56 & 50 & 0.24$^{+0.09}_{-0.08}$ & 0.9994 & 0.80,1.52 & Fit C with +0.03 mag systematic offset\\[+0.5ex]
J & 54 & 57 & 50 & 0.33$^{+0.10}_{-0.09}$ & 0.9912 & 0.72,1.20 & Fit C with $-0.04$ mag systematic offset\\
\\
\multicolumn{8}{l}{\em Clumped Matter Metrics}\\
K & 54 & 57 & 50 & 0.35$^{+0.12}_{-0.10}$ & 0.9984 & 2.90,2.64 & Empty beam metric\tablenotemark{c}\\[+0.5ex]
L & 54 & 56 & 50 & 0.34$^{+0.10}_{-0.09}$ & 0.9974 & 0.94,1.46 & Partially filled beam metric\\

\tablenotetext{a}{0.24 mag intrinsic SNe~Ia luminosity dispersion assumed.}
\vspace{-.1in}
\tablenotetext{b}{Bayesian method of Riess et al. (1996) with
conservative prior (see text and Appendix~A) and 0.10~mag intrinsic
SNe~Ia luminosity dispersion.}
\vspace{-.1in}
\tablenotetext{c}{Assumes additional Bayesian prior of $\Omega_{\rm M} < 3,\Omega_{\Lambda}< 3$.}
\enddata
\end{deluxetable}

\begin{deluxetable}{llcp{3in}}
\scriptsize
\tablenum{4}
\tablecaption{Summary of Uncertainties and Cross-Checks
\label{scorecard}}
\tablehead{& & Uncertainty\tablenotemark{a}\@$\;\;$ on \\
& &($\Omega_{{\rm M}}^{\rm flat}, \Omega_{\Lambda}^{\rm flat})$
= (0.28, 0.72)
}
\startdata
\\
\multicolumn{3}{l}{\bf \hspace{0.86in} 
(A)  Calculated Identified Uncertainties}\\
\multicolumn{2}{l}{\em Statistical Uncertainties (see \S\@5)}&
$\sigma_{\Omega_{{\rm M},\Lambda}^{\rm flat}}=$\\
\hspace{0.2in}& High-redshift SNe  & 0.05  \\
\hspace{0.2in}& Low-redshift SNe   & 0.065 \\
\multicolumn{2}{l}{\bf Total}     & {\bf 0.085}  \\
\\
\\
\multicolumn{3}{l}{\em Systematic Uncertainties from Identified Entities/Processes}\\
&Dust that reddens (see \S\@4.1.2)   & $<$0.03 \\
&\hspace{0.2in} i.e., ${\cal R}_B(z=0.5)\; < \; 2{\cal R}_B({z=0})$ \\

&Malmquist bias difference (see \S\@4.2)&  $<$0.04\\
&K-correction uncertainty (see \S\@2, \S\@3) &  $<$0.025\\
&\hspace{0.2in} including zero-points\\
&Evolution of average (see \S\@4.4)   & $<$0.01\\
&\hspace{0.2in}  SN Ia progenitor metallicity  \\
&\hspace{0.2in} affecting rest-frame $B$ spectral features\\

\multicolumn{2}{l}{\bf Total}   &         {\bf 0.05}\\
\\
\\
\\
\tableline
\\
\multicolumn{3}{l}{\bf \hspace{0.86in} 
(B)  Uncertainties Not Calculated}\\
\multicolumn{3}{l}{\em Proposed/Theoretical Sources of Systematic
Uncertainties}&
{\em Bounds and Tests  (see text)}\\
&Evolving grey dust (see \S\@4.4, \S\@4.1.3) & & Test with $\geq$3-filter color measurements.\\
&\hspace{0.4in} i.e., ${\cal R}_B(z=0.5)\; > \; 2{\cal R}_B({z=0})$ \\
&   \hspace{0.2in}  Clumpy grey dust  (see \S\@5)   &  & Would increase SN mag residual dispersion with $z$.\\
\\
&SN Ia evolution effects (see \S\@4.4)\tablenotemark{b}
&& Test that spectra match on appropriate date for all $z$.\\
&\hspace{0.17in} Shifting distribution of progenitor mass,
& & Compare low- and high-redshift lightcurve rise-times,\\
&\hspace{0.17in} metallicity, C/O ratio & &
and lightcurve timescales before and after maximum. \\
& & & Test width-luminosity relation for low-redshift SNe
across wide range of environments. \\
& & & Compare low- and high-redshift subsets from
 ellipticals/spirals, cores/outskirts, etc. \\

\\
\\
\tableline
\\
\multicolumn{3}{l}{\bf \hspace{0.86in} 
(C)  Cross Checks}\\
\multicolumn{2}{l}{\em Sensitivity to (see \S\@4.5)} &
$\Delta_{\Omega_{{\rm M},\Lambda}^{\rm flat}}=$\\
&Width-luminosity relation & $<$0.03\\
&Non-SN Ia contamination & $<$0.05\\
&Galactic extinction model  & $<$0.04\\
\\
&Gravitational lensing (see \S\@4.3)  & $<$0.06\\
 &by clumped mass \\
\\
\enddata
\tablenotetext{a}{For the redshift distribution of supernovae in this
work, uncertainties in $\Omega_{{\rm M},\Lambda}^{\rm flat}$ correspond
approximately to a factor of 1.3 times uncertainties in the relative supernova
magnitudes.  For ease of comparisons, this table does not distinguish
the small differences between the positive and negative error bars;
see Table~\ref{fitresults} for these.}
\tablenotetext{b}{The comparison of low- and high-redshift lightcurve rise-times discussed
in Section 4.4 theoretically limits evolutionary changes in the
zero-point of the lightcurve width-luminosity relation to less than
$\sim$0.1 mag, i.e. $\Delta_{\Omega_{{\rm M},\Lambda}^{\rm flat}} \lesssim
0.13$.
 }
\end{deluxetable}

\clearpage

\begin{figure}[p]
\psfig{file=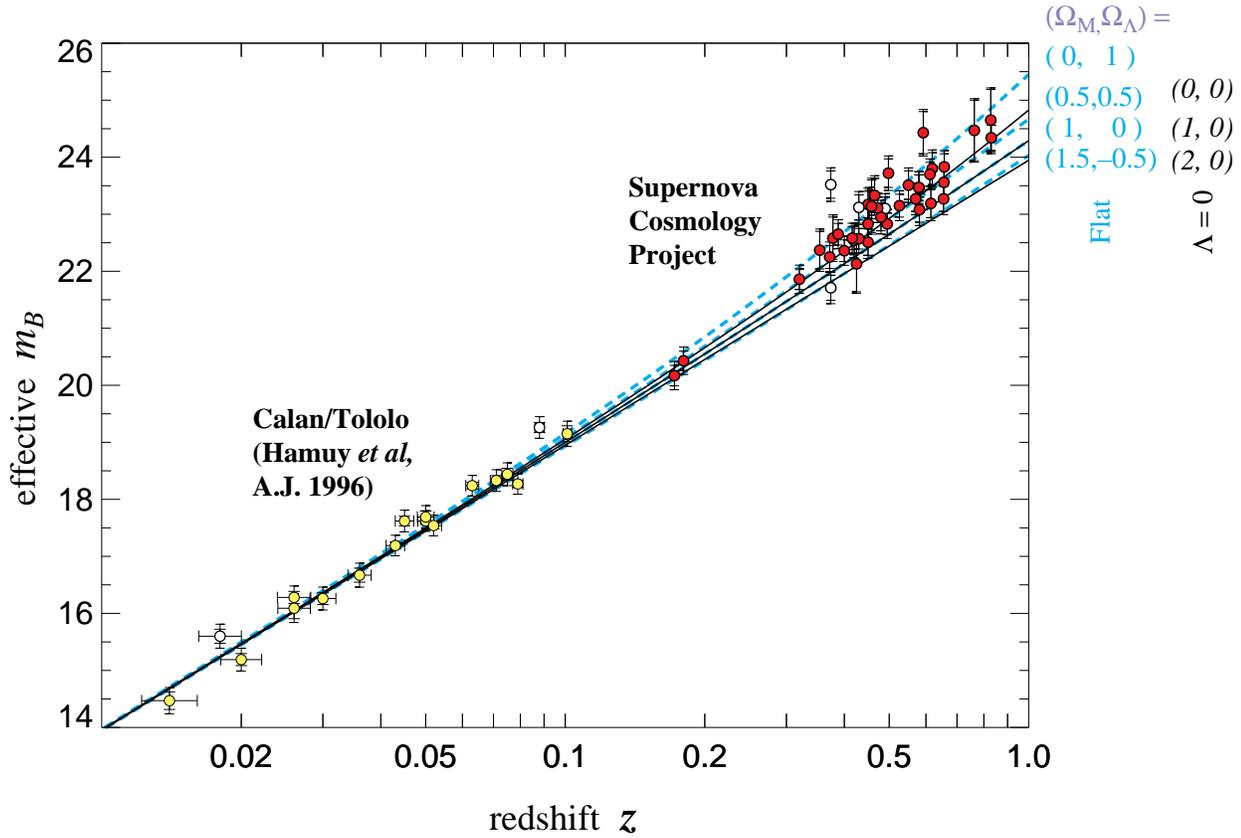,height=4.8in}
\caption{
Hubble diagram for 42
high-redshift Type Ia supernovae from the Supernova Cosmology Project,
and 18 low-redshift Type Ia supernovae from the Cal\'{a}n/Tololo Supernova
Survey, after correcting both sets for the SN~Ia lightcurve width-luminosity
relation. The inner error bars show the uncertainty due to measurement
errors, while the outer error bars show the total uncertainty when
the intrinsic luminosity dispersion, 0.17 mag, of lightcurve-width-corrected
Type Ia supernovae is added in quadrature.
The unfilled circles indicate supernovae not included in Fit~C.
The horizontal error bars represent the assigned
peculiar velocity uncertainty of 300 km s$^{-1}$.
The solid curves are the theoretical $m_B^{\rm effective}(z)$
for a range of cosmological models with zero cosmological constant:
$(\Omega_{\rm M},\Omega_\Lambda) = (0,0)$ on top, $(1,0)$ in middle
and (2,0) on bottom.  The dashed curves are for a range of flat
cosmological models: $(\Omega_{\rm M},\Omega_\Lambda) = (0,1)$
on top, $(0.5,0.5)$ second from top, $(1,0)$ third from top,
and (1.5,-0.5) on bottom.
\label{hubdiaglog}}
\end{figure}
\clearpage

\begin{figure}[p]
\psfig{file=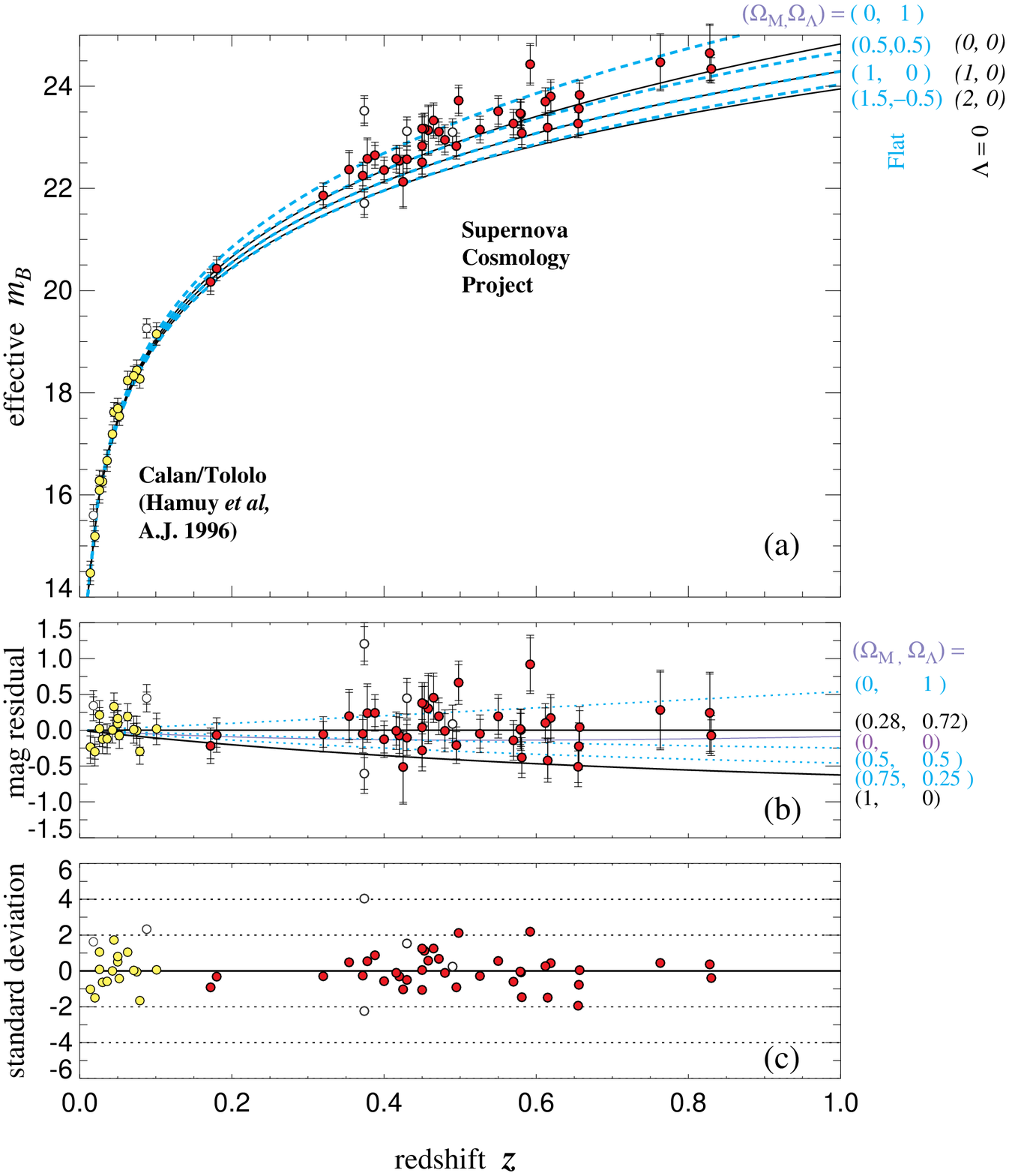,height=8.8in}
\caption{ \label{hubdiag}}
\end{figure}
\clearpage
(a) Hubble diagram for 42
high-redshift Type Ia supernovae from the Supernova Cosmology Project,
and 18 low-redshift Type Ia supernovae from the Cal\'{a}n/Tololo Supernova
Survey, plotted on a linear redshift scale to display details at
high redshift.   The symbols and curves are as in Figure 1.
(b)  The magnitude residuals from the best-fit flat cosmology
for the Fit~C supernova subset,
$(\Omega_{\rm M},\Omega_\Lambda) = (0.28,0.72)$.
The dashed curves are for a range of flat
cosmological models: $(\Omega_{\rm M},\Omega_\Lambda) = (0,1)$
on top, $(0.5,0.5)$ third from bottom, $(0.75,0.25)$ second from bottom,
and (1,0) is the solid curve on bottom.  The middle solid curve is for
$(\Omega_{\rm M},\Omega_\Lambda) = (0,0)$.
Note that this plot is practically identical to
the magnitude residual plot for the best-fit unconstrained
cosmology of Fit~C, with $(\Omega_{\rm M},\Omega_\Lambda) = (0.73,1.32)$.
(c)  The uncertainty-normalized
residuals from the best-fit flat cosmology
for the Fit~C supernova subset,
$(\Omega_{\rm M},\Omega_\Lambda) = (0.28,0.72)$.

\clearpage
\begin{figure}[p]
\psfig{file=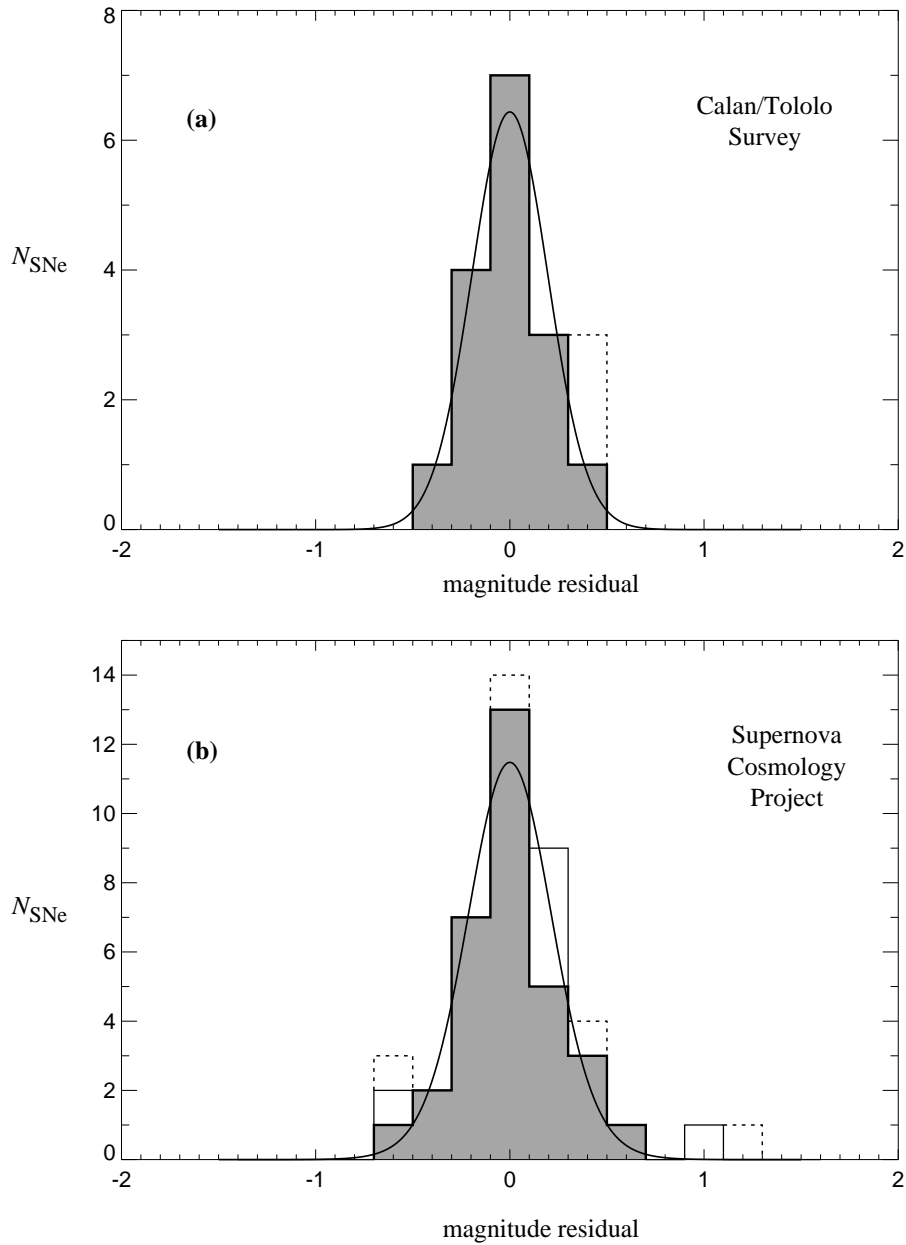,height=6.5in}
\caption{
The distribution of  restframe $B$-band magnitude residuals from the best-fit
flat cosmology for the Fit~C supernova subset,
for (a) 18 Cal\'{a}n/Tololo supernovae, at redshifts $z \le0.1$
and (b) 42 supernovae from the Supernova Cosmology Project,
at redshifts between 0.18 and 0.83.  The darker shading indicates those
residuals with uncertainties less than
0.35 mag, unshaded boxes indicate uncertainties greater than 0.35 mag, and
dashed boxes indicate the supernovae that are
excluded from Fit C.
The curves show the expected magnitude residual
distributions if they are drawn from normal distributions
given the measurement uncertainties and 0.17 mag of intrinsic
SN~Ia dispersion.  The low-redshift expected distribution matches
a Gaussian with $\sigma = 0.20$ mag (with error on the mean of
0.05 mag), while the high-redshift expected
distribution matches a Gaussian with $\sigma = 0.22$ mag (with error on the
mean of 0.04 mag).
\label{residhists}}
\end{figure}

\clearpage

\begin{figure}[p]
\psfig{file=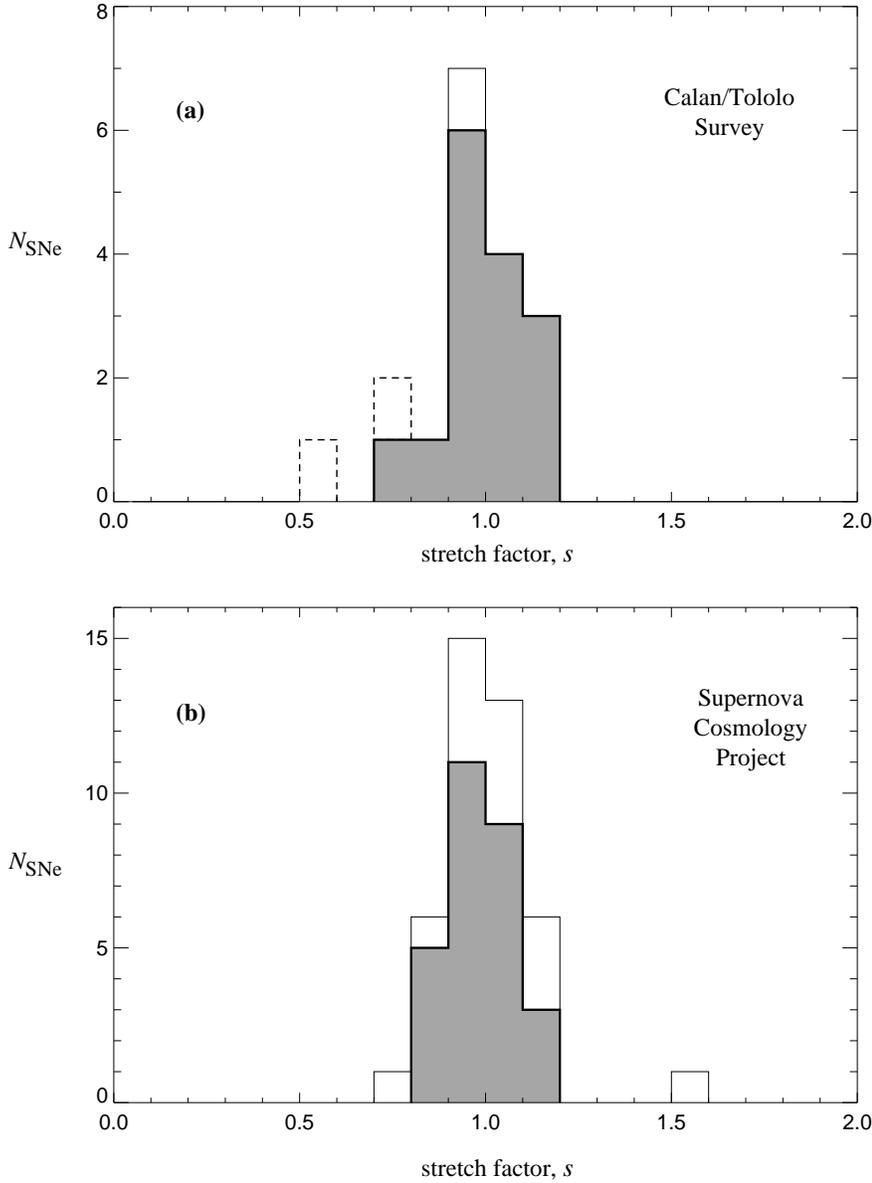,height=6.2in}
\caption{
The distribution of lightcurve widths for (a) 18
Cal\'{a}n/Tololo supernovae, at redshifts $z \le0.1$
and (b) 42 supernovae from the Supernova Cosmology Project,
at redshifts between 0.18 and 0.83.  The lightcurve
widths are characterized
by the ``stretch factor,'' $s$, that stretches or contracts the time
axis of a template SN~Ia lightcurve to best fit the observed lightcurve
for each supernova
\protect\cite[see 
Perlmutter et al. 1995a, 1997e;][]{kim_stretch98, gold_dilate98}.
The template has been time-dilated by a factor $1+z$ before fitting to the
observed lightcurves to account
for the cosmological lengthening of the supernova timescale
\protect\cite[]{goldhaberaigua,leibundgutdilation}.
The shading indicates those measurements of $s$ with uncertainties
less than 0.1, and the dashed lines indicate the two supernovae
that are removed from the fits after Fit~A.   These two excluded
supernovae are the most significant
deviations from $s=1$
(the highest-stretch supernova in panel (b)
has an uncertainty of $\pm$0.23 and hence is not a significant
outlier from $s=1$); the remaining low- and
high-redshift distributions have
almost exactly the same error-weighted means:
$\langle s \rangle_{\rm Hamuy} = 0.99 \pm 0.01$
and $\langle s \rangle_{\rm SCP} = 1.00 \pm 0.01$.
\label{stretchhists}}
\end{figure}

\clearpage

\begin{figure}[p]
\psfig{file=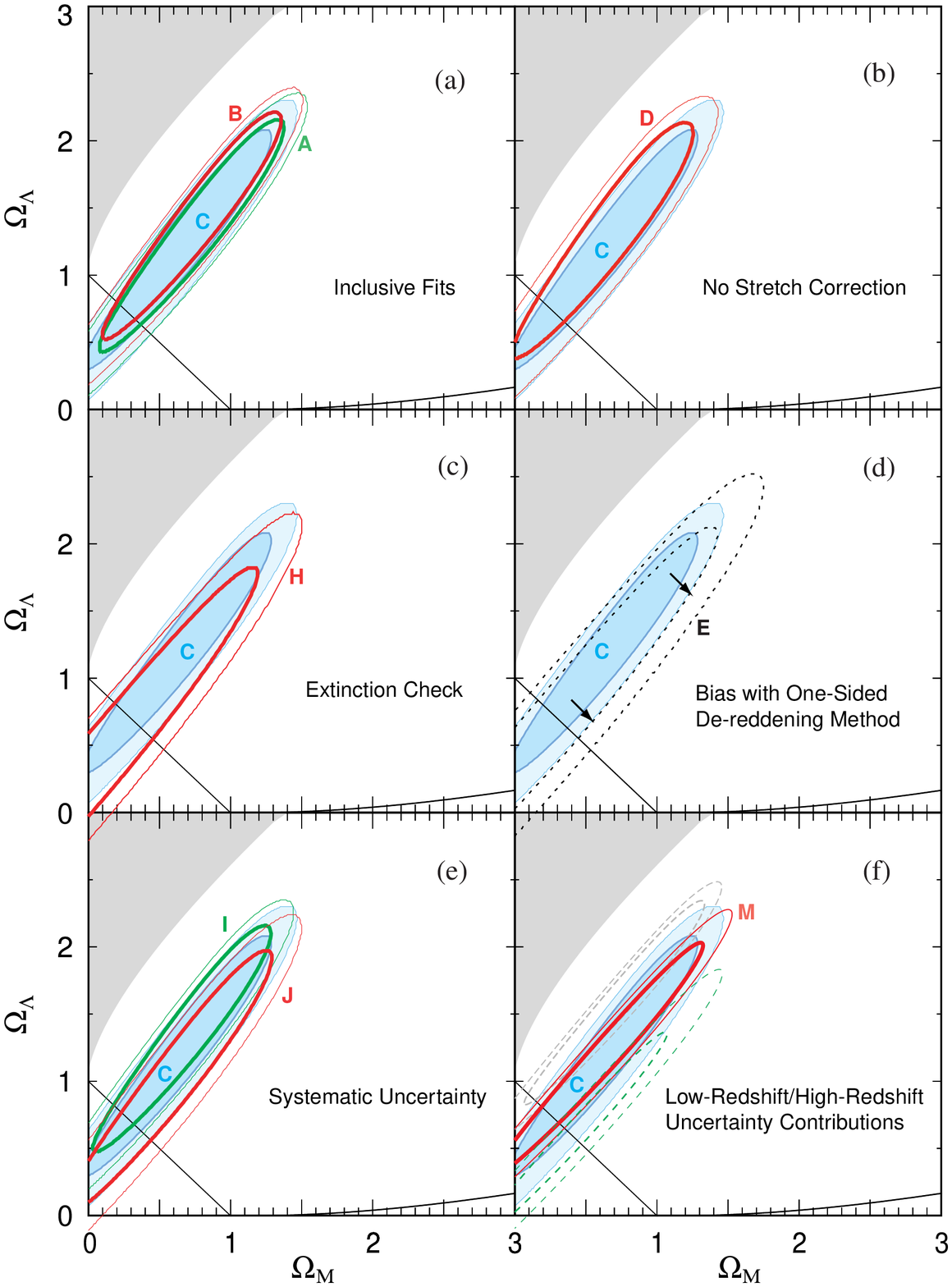,height=8.4in}
\caption{ \label{confmulti}}
\end{figure}
\clearpage
Comparison of best-fit confidence regions in the
$\Omega_{\rm M}$--$\Omega_\Lambda$ plane.  Each subpanel shows
the result of Fit~C (shaded regions)
compared with fits to different subsets of
supernovae, or variant analyses for the same subset of supernovae,
to test the robustness of the Fit~C result.  Unless otherwise indicated,
the 68\% and 90\% confidence regions in the
$\Omega_{\rm M}$--$\Omega_\Lambda$ plane are shown, after integrating
the four-dimensional fits
over the other two variables, ${\cal M}_B$ and $\alpha$.
The ``no-big-bang'' upper-left shaded region,
the flat-universe diagonal line, and the infinite expansion line are
shown as in Figure~\ref{40conf} for ease of comparison.  The subpanels
show: (a) Fit~A of all 60 supernovae;
and Fit~B of 56 supernovae, excluding the two outliers
from the lightcurve width distribution and the two remaining
statistical outliers.  Fit~C further excludes the two likely reddened
high-redshift supernovae.
(b) Fit~D of the same 54-supernova subset as in Fit~C, but with no
correction for the lightcurve width-luminosity relation.
(c) Fit~H of the subset of supernovae with color measurements,
after excluding the reddest 25\% (9 high-redshift supernovae)
and the two faint high-redshift supernovae with large uncertainties
in their color measurements.  The close match
to the confidence regions of Fit~C indicates that any extinction
of these supernovae is quite small, and
not significant in the fits of the cosmological parameters.
(d) The 68\% confidence region for
Fit~E of the 53 supernovae with color measurements
from the Fit~B dataset, but following the Bayesian reddening-correction
method of \protect\cite{rpk96}.  This method, when used with any
reasonably conservative prior (i.e., somewhat broader than
the likely true extinction distribution; see text), can produce a result
that is biased, with an approximate bias direction and worst-case
amount indicated by the arrows.
(e) Fits~I and J are identical to Fit~C, but with 0.03 or 0.04 magnitudes
added or subtracted from each of the high-redshift supernova measurements,
to account for the full range of identified 
systematic uncertainty in each direction.
Other hypothetical sources of systematic uncertainty 
(see Table~\ref{scorecard}B) are not included.
(f) Fit~M is a separate two-parameter ($\Omega_{\rm M}$ and $\Omega_\Lambda$)
fit of just the high-redshift supernovae, using the values of
${\cal M}_B$ and $\alpha$ found from the low redshift supernovae.  The
dashed confidence regions show the approximate range of uncertainty
from these two low-redshift-derived parameters, added to the systematic
errors of Fit~J.  Future well-observed low-redshift supernovae can constrain
this dashed-line range of uncertainty.

\clearpage

\begin{figure}[p]
\psfig{file=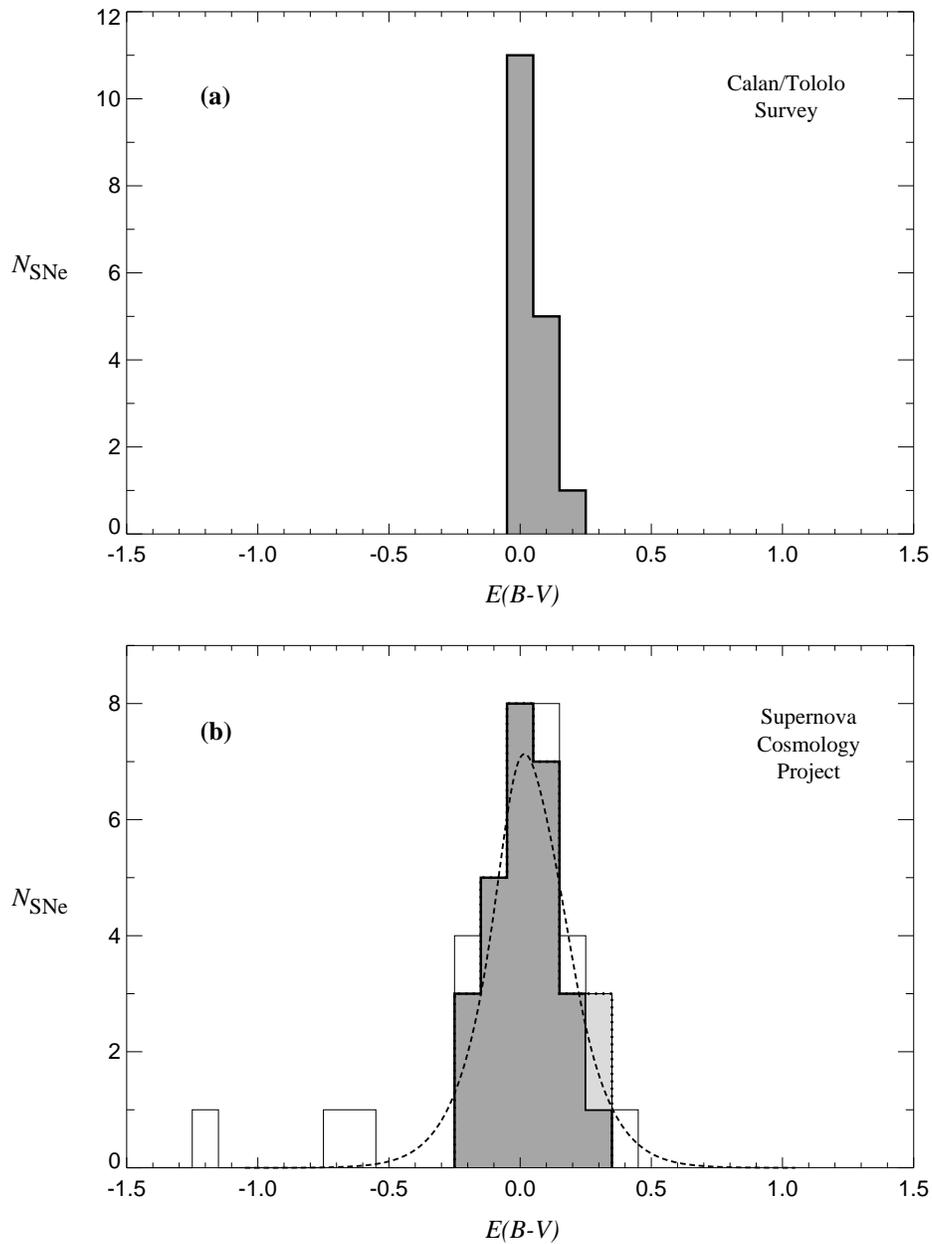,height=6.5in}
\caption{
(a) The restframe $B\!-\!V$ color excess distribution
for 17 of 18 Cal\'{a}n/Tololo supernovae
(see text), corrected for Galactic extinction
using values from Schlegel {\em et. al.} 1998.
(b) The restframe $B\!-\!V$ color excess for the
36 high-redshift supernovae for which restframe $B\!-\!V$ colors were
measured, also corrected for Galactic extinction.
The darker shading indicates those $E(B\!-\!V)$ measurements
with uncertainties less than
0.3 mag, unshaded boxes indicate uncertainties greater than 0.3 mag, and
the light shading indicates the two supernovae that are
likely to be reddened based on their joint probability in color excess
and magnitude residual from Fit B.
The dashed curve shows the expected high-redshift $E(B\!-\!V)$
distribution if the low-redshift distribution
had the measurement uncertainties of the high-redshift supernovae
indicated by the dark shading.
Note that most of the color-excess dispersion for the high-redshift
supernovae is due to the rest-frame $V$-band measurement
uncertainties, since the rest-frame $B$-band uncertainties are
usually smaller.
\label{bminv}}
\end{figure}

\clearpage

\begin{figure}[p]
\psfig{file=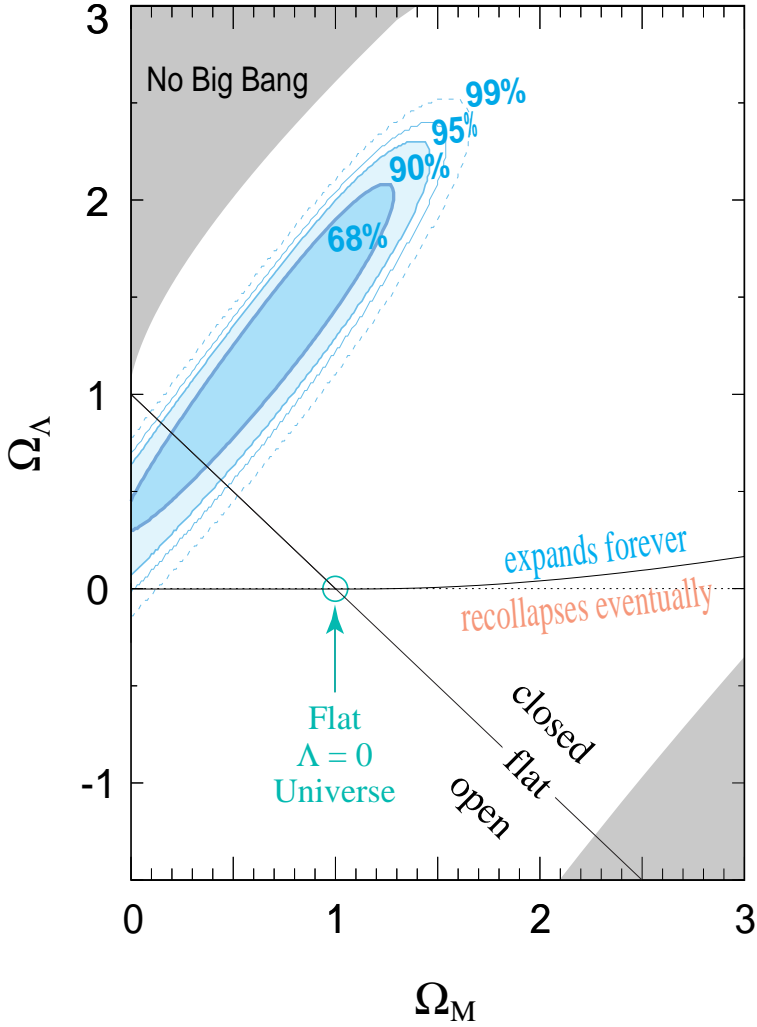,height=5.7in}
\caption{
Best-fit confidence regions in the
$\Omega_{\rm M}$--$\Omega_\Lambda$ plane for our primary analysis,
Fit~C.  The 68\%, 90\%, 95\%, and 99\% statistical confidence regions in the
$\Omega_{\rm M}$--$\Omega_\Lambda$ plane are shown, after integrating the
four-dimensional fit over ${\cal M}_B$ and $\alpha$.
(The table of this two-dimensional
probability distribution is available at http://www-supernova.lbl.gov/.)
See Figure~\ref{confmulti}(e) for limits on the small 
shifts in these contours due to identified systematic uncertainties.
Note that the spatial curvature of the universe---open, flat,
or closed---is not determinative of the future of the universe's expansion,
indicated by the near-horizontal solid line.
In cosmologies
above this near-horizontal line the
universe will expand forever, while below this line the
expansion of the universe will eventually come to a halt and
recollapse.  This line is not quite horizontal because
at very high mass density there is a region
where the mass density can bring the expansion to a halt before the
scale of the universe is big enough that the mass density is dilute
with respect to the cosmological constant energy density.
The upper-left shaded region, labeled
``no big bang,'' represents ``bouncing universe'' cosmologies with
no big bang in the past \protect\cite[see][]{carrollpressturner}.  The lower
right shaded region corresponds to a universe that is younger than
the oldest heavy elements \protect\cite[]{schramm90},
for any value of $H_0 \ge 50$ km s$^{-1}$ Mpc$^{-1}$.
\label{40conf}}
\end{figure}

\clearpage

\begin{figure}[p]
\begin{center}
\psfig{file=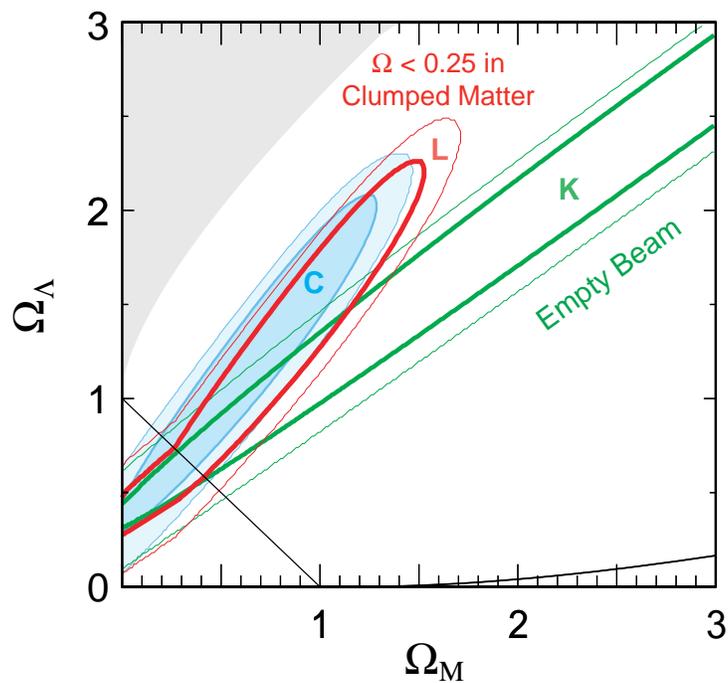,height=4.2in}
\end{center}
\caption{
Best-fit 68\% and 90\% confidence regions in the
$\Omega_{\rm M}$--$\Omega_\Lambda$ plane for
cosmological models with small scale clumping of matter
(e.g., in the form of MACHOs) compared with the
Friedman-Robertson-Walker model of Fit~C, with
smooth small-scale matter distribution.
The shaded contours (Fit~C) are the confidence regions fit to a
Friedman-Robertson-Walker magnitude-redshift relation.
The extended confidence strips (Fit~K) are for a fit
of the Fit~C supernova set
to an ``empty beam'' cosmology, using the ``partially filled beam''
magnitude-redshift relation with a filling factor $\eta = 0$,
representing an extreme case in which all mass is in compact
objects.  The Fit~L un-shaded contours represent a somewhat
more realistic partially-filled-beam fit, with clumped matter ($\eta = 0$)
only accounting for up to $\Omega_{\rm M} = 0.25$ of the
critical mass density, and any matter beyond that amount smoothly
distributed (i.e., $\eta$ rising to 0.75 at $\Omega_{\rm M} = 1$).
\label{40confeta}}
\end{figure}

\clearpage

\begin{figure}[p]
\begin{center}
\psfig{file=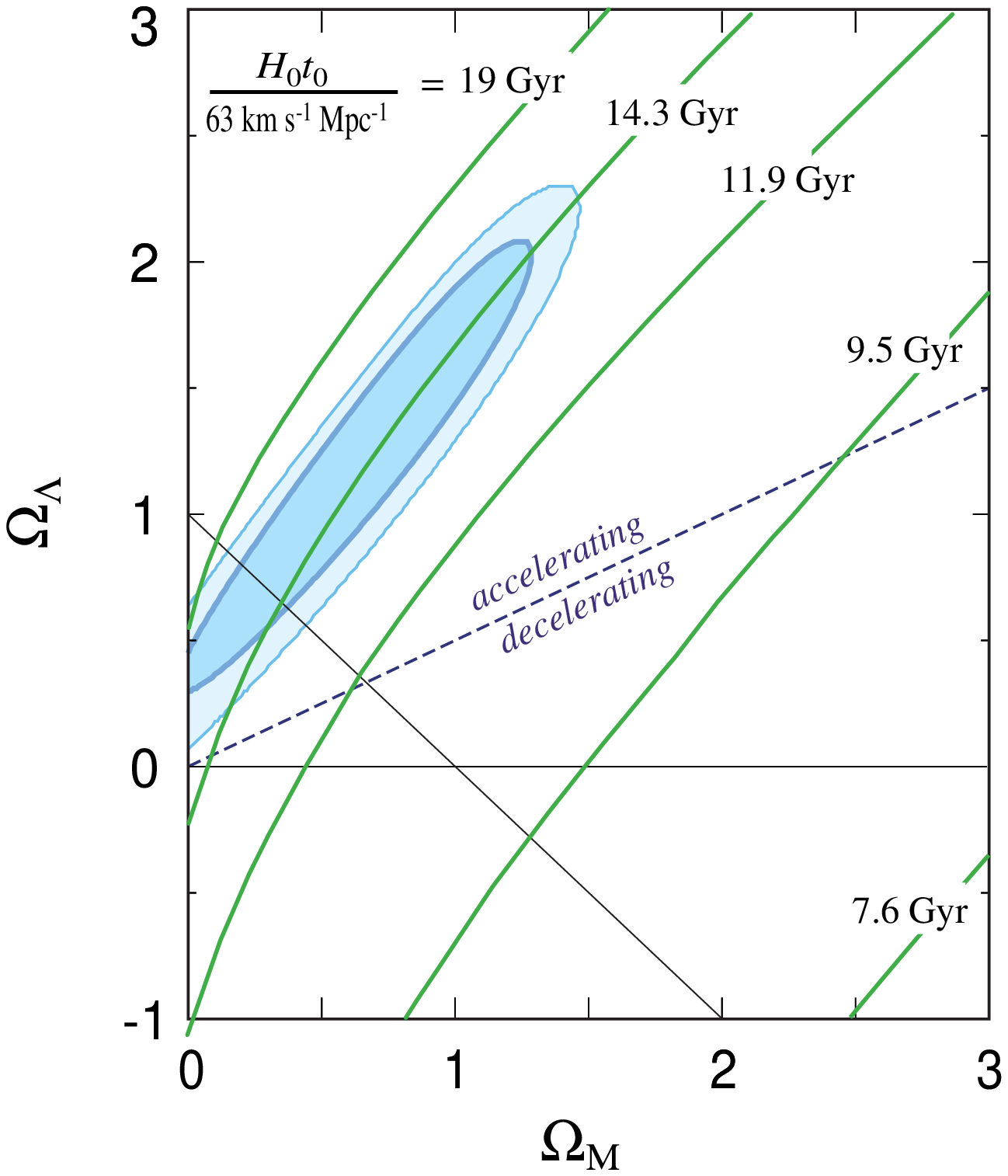,height=6.5in}
\end{center}
\caption{
Isochrones of constant $H_0 t_0$, the
age of the universe relative to the Hubble time,
$H_0^{-1}$, with the best-fit 68\% and 90\% confidence regions in the
$\Omega_{\rm M}$--$\Omega_\Lambda$ plane for the primary analysis,
Fit~C.  The isochrones are labeled for the case of $H_0=63$ km s$^{-1}$
Mpc$^{-1}$, representing a typical value found from studies of
SNe~Ia \protect\cite[]{hametal96,rpk96,sahaetal97,tripp98}.
If $H_0$ were taken to be 10\% larger
\protect\cite[i.e., closer to the values in][]{freedmaniau97},
the age labels would be 10\% smaller.
The diagonal line labeled accelerating/decelerating is drawn for $q_0 \equiv
\Omega_{\rm M}/2 - \Omega_\Lambda = 0$, and divides the cosmological
models with an accelerating or decelerating expansion at the present time.
\label{ages}}
\end{figure}
\clearpage

\begin{figure}[p]
\begin{center}
\psfig{file=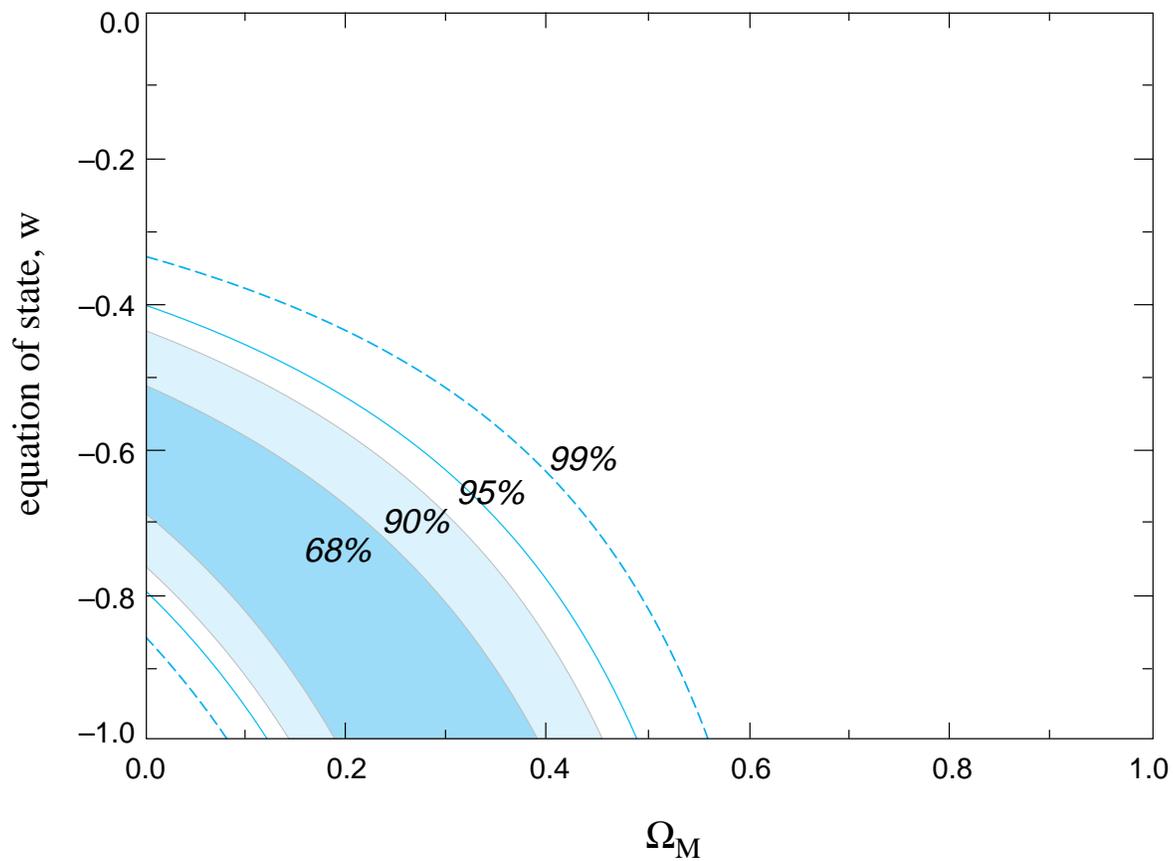,height=4.5in}
\end{center}
\caption{
Best-fit 68\%, 90\%, 95\%, and 99\% confidence regions in the
$\Omega_{\rm M}$--$w$ plane for an additional energy density
component, $\Omega_w$, characterized by an equation-of-state $w = p/\rho$.
(If this energy density component is Einstein's cosmological constant,
$\Lambda$, then the equation of state is $w = p_\Lambda/\rho_\Lambda = -1$.)
The fit is for the supernova subset of our primary analysis,
Fit~C, constrained to a flat cosmology ($\Omega_{\rm M} + \Omega_w =1$).
The two variables ${\cal M}_B$ and $\alpha$ are included in the fit,
and then integrated over to obtain the two-dimensional
probability distribution shown.
\label{wconf}}
\end{figure}

\end{document}